\documentclass{jfm}
\usepackage{graphicx}
\usepackage{natbib}
\usepackage{color}
\usepackage{pifont}
\usepackage{array,multirow}
\usepackage{scrtime}
\usepackage{float}
\usepackage{overpic,epstopdf,rotating,dashrule}
\usepackage{amsmath,amscd,amssymb}
\usepackage{booktabs}
\usepackage{siunitx}
\usepackage{commath}
\usepackage{subfigure}
\renewcommand{\thesubfigure}{\thefigure.\arabic{subfigure}}
\makeatletter
\renewcommand{\p@subfigure}{}
\renewcommand{\@thesubfigure}{\thesubfigure)\hskip\subfiglabelskip}
\makeatother
\usepackage{csquotes}

\ifCUPmtlplainloaded \else
  \checkfont{eurm10}
  \iffontfound
    \IfFileExists{upmath.sty}
      {\typeout{^^JFound AMS Euler Roman fonts on the system,
                   using the 'upmath' package.^^J}%
       \usepackage{upmath}}
      {\typeout{^^JFound AMS Euler Roman fonts on the system, but you
                   dont seem to have the}%
       \typeout{'upmath' package installed. JFM.cls can take advantage
                 of these fonts,^^Jif you use 'upmath' package.^^J}%
      }
  \else
  \fi
\fi

\ifCUPmtlplainloaded \else
  \checkfont{msam10}
  \iffontfound
    \IfFileExists{amssymb.sty}
      {\typeout{^^JFound AMS Symbol fonts on the system, using the
                'amssymb' package.^^J}%
       \usepackage{amssymb}%
         \let\leq=\leqslant
         
      }{}
  \fi
\fi

\ifCUPmtlplainloaded \else
  \IfFileExists{amsbsy.sty}
    {\typeout{^^JFound the 'amsbsy' package on the system, using it.^^J}%
     \usepackage{amsbsy}}
    {}
\fi

\newcommand\Rey{\mbox{\textit{Re}}}  % Reynolds number
\newsavebox{\astrutbox}
\sbox{\astrutbox}{\rule[-5pt]{0pt}{20pt}}

\title[Closed-loop control using machine learning control]{Closed-loop control of an experimental mixing layer using machine learning control}
\author[V. Parezanovi{\'c} and friends]%
{Vladimir Parezanovi{\'c}$^1$
  \thanks{Email address for correspondence: vladimir.parezanovic@univ-poitiers.fr},\ns
Thomas Duriez$^{1,2}$,\ns
Laurent Cordier$^1$,\break
Bernd R. Noack$^1$,\ns
Jo{\"e}l Delville$^1$,\ns
Jean-Paul Bonnet$^1$,\break
Marc Segond$^3$,\ns
Markus Abel$^{3,4,5}$\ns
and\ns
Steven L. Brunton$^{6}$
}
\affiliation{$^1$Institut PPRIME, CNRS - {Universit\a'e de Poitiers} - ENSMA, UPR 3346, D\a'epartement Fluides, Thermique, Combustion, CEAT, 43, rue de l'A\a'erodrome, F-86036 Poitiers Cedex, France\\[\affilskip]
$^2$Laboratorio de FluidoDinamica - Facultad de Ingeneria CONICET - Universidad de Buenos Aires, Paseo Colon 850, Ciudad Autonoma de Buenos Aires, Argentina\\[\affilskip]
$^3$Ambrosys GmbH, Albert-Einstein-Str.\ 1-5, D-14469 Potsdam, Germany\\[\affilskip]
$^4$LEMTA, 2 avenue de la For\a^et de Haye, F-54518 Vandoeuvre-l\a`es-Nancy Cedex, France\\[\affilskip]
$^5$University of Potsdam, Karl-Liebknecht-Str. 24/25, D-14476 Potsdam, Germany\\[\affilskip]
$^6$University of Washington, Applied Math Department, Seattle, WA 98195, USA}
\date{?; revised ?; accepted ?. - To be entered by editorial office}
\begin{document}

\maketitle
% =============== ABSTRACT ===========================================%
\begin{abstract}
A novel framework for closed-loop control of turbulent flows is tested in an experimental mixing layer flow. This framework, called Machine Learning Control (MLC), provides a model-free method of searching for the best function, to be used as a control law in closed-loop flow control. MLC is based on genetic programming, a function optimization method of machine learning. In this article, MLC is benchmarked against classical open-loop actuation of the mixing layer. Results show that this method is capable of producing sensor-based control laws which can rival or surpass the best open-loop forcing, and be robust to changing flow conditions. Additionally, MLC can detect non-linear mechanisms present in the controlled plant, and exploit them to find a better type of actuation than the best periodic forcing.    
\end{abstract}
% =============== INTRODUCTION =======================================%
\section{Introduction}\label{intro}

Closed-loop turbulence control is fast gaining in significance as a research topic in fluid mechanics. Engineering benefits of such control applied to a realistic flow can be enormous. Some examples include: reduction of $CO_2$ emissions and fuel consumption optimization for large transport vehicles through drag reduction, optimization of green energy harnessing from wind and water turbines by continuous control of lift and drag properties of turbine blades, countless opportunities for medical applications, etc. Among all the control strategies, feedback control has been extremely successful in fluid dynamics. For many applications, such as the stabilization of unstable steady-states, linear model-based feedback control suppresses instabilities and keeps the flow approximately linear so that models and controllers remain effective. Some examples include the transitional channel flow~\citep{ilak:2008,Ilak2009}, flows of backward-facing step~\citep{Herve2012jfm}, the flat plate boundary layer~\citep{bagheri:2009,Semeraro2013}, and cavity-flow oscillations~\citep{Rowley2002,illingworth:2010}, although there are many others. Balanced reduced-order models have been especially effective because they capture the most relevant flow states that are both controllable and observable, based on input-output data from simulations or experiments~\citep{rowley:05pod,ERA:2009}.

For fully turbulent flows, however, most controllers rely on either open-loop or model-free adaptive control strategies. This is largely due to the fact that turbulent flows are characterized by strongly nonlinear dynamics that evolve on a high-dimensional attractor. For these flows, linear models are incapable of capturing many significant mechanisms, including broadband frequency cross-talk. Some successful studies of wake stabilization with high-frequency actuation \citep{Glezer2005aiaaj,Thiria2006jfm,Luchtenburg2009jfm} and low-frequency forcing \citep{Pastoor2008jfm} show that frequency cross-talk can be a crucial potential target for actuation. Moreover, the goal of control in a turbulent system may not be the stabilization of a laminar solution, but rather the enhancement of turbulent energy. In the appendix, we investigate the performance of balanced linear models of the mixing layer experiment and illustrate the severe limitations.  

The adaptive control approaches are mostly based on a slow variation of amplitude and frequency of periodic actuation until continuously monitored performance reaches a maximum~\citep{King2010book}. Examples of such approaches include resonance frequency adaptation using extremum seeking \citep{beaudoin2006bluff}, and amplitude selection using slope seeking \citep{King2006ieeemed}. In these cases, the response of non-linear mechanisms to actuation is taken into account in performance evaluation, but the periodic forcing is, by nature, incapable of in-time targeting of specific flow events, in order to affect these non-linearities in-time directly.  

A few successful examples of an in-time control are the reduction of skin friction in wall turbulence~\citep{Choi1994jfm} and phasor control for turbulence with a dominant oscillatory structure~\citep{Samimy2007jfm}. In the former, a simple opposition control in the viscous sublayer is already effective; while in the latter, the control design requires a robust phase detection from the sensors and effective gain scheduling for the actuators. In general, a robust control-oriented reduced-order model, which could at least resolve the turbulent coherent structures and transients between them, would be required for an experimental application of a model-based control design.

Aerodynamics turns its eye once more to bird and insect flight for inspiration. A bird is a flying object whose surface is completely covered with sensors, and at the same time fully elastic so that each small part acts as an independent actuator. Seeing a bird in flight, one might think this action is effortless, but humans must think in terms of complex transfer functions between millions of sensors (skin receptors) and thousands of actuators (feathers). Numerical exploration of bio-inspired actuation already yields intriguing results~\citep{bagheri2012spontaneous}. Development of modern actuation systems, such as micro-valves, piezoelectric, synthetic jets, plasmas, etc.~\citep{cattafesta2011rev,corke2010plasma,glezer2002revjets}, provides progressively increasing options for application of control in experimental studies. Concurrent increase in capabilities of real-time systems with large input/output data rates and computational power bring us ever closer to having an affordable, yet powerful and compact multiple-input-multiple-output (MIMO) control system. Could we also take inspiration from the living organisms in learning how to create control laws for such complex sensor/actuator systems, in flow conditions where linear models and adaptive control do not suffice?

Recently, we have proposed Machine Learning Control (MLC) as a framework for a model-free closed-loop control of turbulent flows~\citep{duriez_jfm_2014,Parezanovic_FTC_2014,gautier2014jfm}. MLC is based on genetic programming~\citep{koza1992genetic,koza1999genetic}, a biologically-inspired machine learning function optimization method, used to design controllers in robotics~\citep{wahde2008biologically,lewis1992genetic,nordin1997line}. The use of machine learning for control~\citep{fleming2002evolutionary} also includes genetic algorithms which can only be used to optimize control parameters~\citep{de2014optimizing} and artificial neural networks~\citep{noriega1998direct}. 

The current article will present the application of MLC in a turbulent mixing layer. Open-loop forcing of a shear layer has been proven effective at changing its stream-wise evolution~\citep{Oster1982,Fiedler_1998_Cargese,Wiltse_GlezerJFM_1993}. In this experiment, actuation is applied at the trailing edge of the splitter plate with a goal of changing the local properties of the flow, such as the fluctuating energy content or thickness of the mixing layer. The sensors placed at the location of interest, downstream of the trailing edge, are subject to a convective time delay. MLC is employed in finding the best sensor-based, feedback control law, which can rival or exceed the efficiency of open-loop forcing, while ensuring robustness to variations in the mean flow conditions. 

The main features of the experimental installation and the sensor/actuator system are described in \S~\ref{sec:setup}. Formulation of the objective functions, a detailed description of the methodology of MLC and its practical application in the experiment are described in \S~\ref{sec:strategy}. In \S~\ref{sec:results} results of the un-actuated flow are presented, followed by a study of the mixing layer's response to open-loop forcing, in order to have a base of comparison with closed-loop results. The performance of the best closed-loop control laws, obtained by MLC, is then presented for maximization and minimization of objective functions in two different mixing layer speed configurations. An experiment using high-gain actuation amplitude control, and the implications of its effects on the MLC process, is also discussed. In \S~\ref{sec:discussion}, we examine different parameters and the convergence process of experimental MLC, as well as the robustness of open- and closed-loop control to varying flow conditions. The results are summarized and a discussion on MLC effectiveness and future development is provided in \S~\ref{sec:conclusion}. In Appendix~\ref{sec:era}, eigensystem realization algorithm (ERA) and observer/Kalman filter identification (OKID) are performed on the measurements from the mixing layer. The results of this procedure are presented and discussed as our model-based benchmark.  
% =============== EXPERIMENTAL SETUP =================================%
\section{Experimental setup}\label{sec:setup}

In this section, the wind-tunnel facility and the geometry of the key elements in the test section (splitter plate) are described. We present the design of the actuator system and discuss the control of the actuation amplitude and evaluation of the invested energy of actuation. Finally, sensor and acquisition systems are described.   

\subsection{Wind tunnel}\label{sec:wind}

\begin{figure*}
  \centerline{\includegraphics[width=1\textwidth]{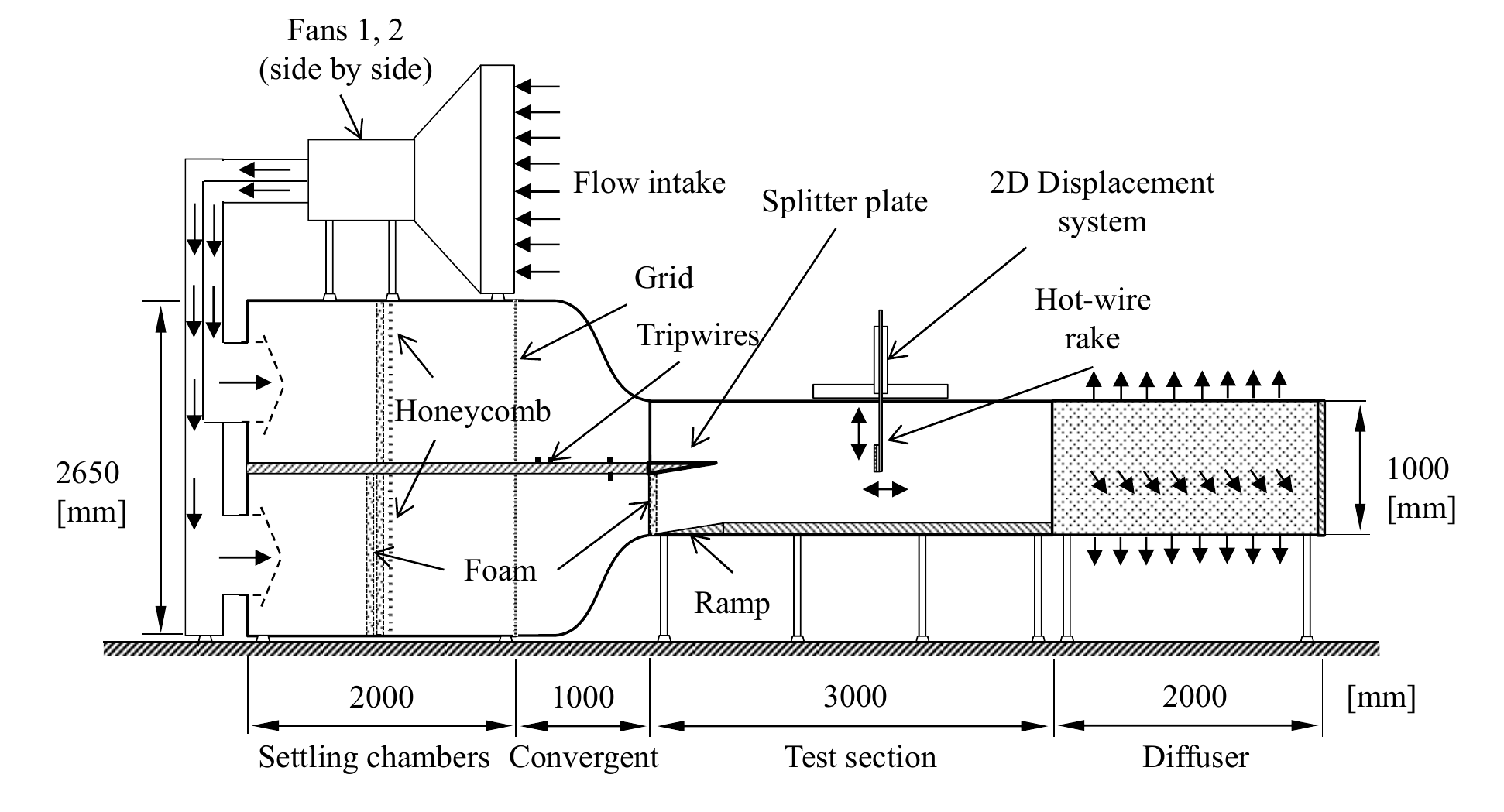}}% Images in 100% size
  \caption{Dual stream TUCOROM wind tunnel facility schematic.}
\label{fig:experiment}
\end{figure*}

The wind tunnel facility has been designed and built as a part of the TUCOROM\footnote{ANR project: "TUrbulence COntrol using Reduced-Order Models"} project. It features a twin-turbine design with two separate air streams, running in parallel and meeting in the test section to produce a mixing layer. The turbines are capable of propelling the air at velocities up to \SI{12}{\meter\per\second} at the inlet of the test section. These velocities are however not practically obtained due to the installation of foam layers at key points along the stream paths (as shown in figure~\ref{fig:experiment}) whose function is to homogenize the flow and prevent the forming of large structures upstream of the test section. This was a necessary design choice as the initial stages of the wind tunnel must be compact, in order for the test section to be long enough to allow for unimpeded spatial development of the mixing layer flow. In addition, the convergent part of the wind tunnel has trip-wires installed which ensure that the boundary layers at the end of the splitter plate are turbulent for operation at higher speeds \textit{i.e.} when a fully turbulent mixing layer is desired.   

The test section is $l_x \times l_y \times l_z = \SI{3}{\meter} \times \SI{1}{\meter} \times \SI{1}{\meter}$, with a square cross-section. It begins at the throat of the convergent part of the wind tunnel, where the surface separating the two air streams terminates with a splitter plate. The splitter plate maintains a horizontal upper surface (from where the upper stream originates), while its lower surface is angled at 8 degrees so as to connect with the upper surface and form a relatively thin trailing edge (see figure~\ref{fig:actuator}). The initial thickness of the splitter plate is \SI{8}{\centi\meter} and the trailing edge is only \SI{2}{\milli\meter} thick. The splitter plate spans the entire width of the test section inlet, and extends by \SI{39}{\centi\meter} into the test section. The bottom of the test section has been raised by inserting a ramp of the same shape as the lower surface of the splitter plate, in order to compensate for adverse pressure gradient on the lower stream boundary layer caused by the angling of the splitter plate. This reduces the height of the test section by \SI{8}{\centi\meter}. An additional foam layer is inserted at the convergent throat just upstream of the angled splitter plate in order to further stabilize the lower stream and its boundary layer. All these provisions limit the maximum operational speeds to \SI{9.5}{\meter\per\second} for the upper and \SI{1.7}{\meter\per\second} for the lower stream. 

The test section terminates with a $\SI{2}{\meter} \times \SI{1}{\meter} \times \SI{1}{\meter}$ foam-lined diffuser which recycles the air into the surrounding environment and prevents the structures created by the mixing layer from returning into the turbine inlets.       

\subsection{Actuator system}\label{sec:actuators}

\begin{figure*}
  \centerline{\includegraphics[width=1\textwidth]{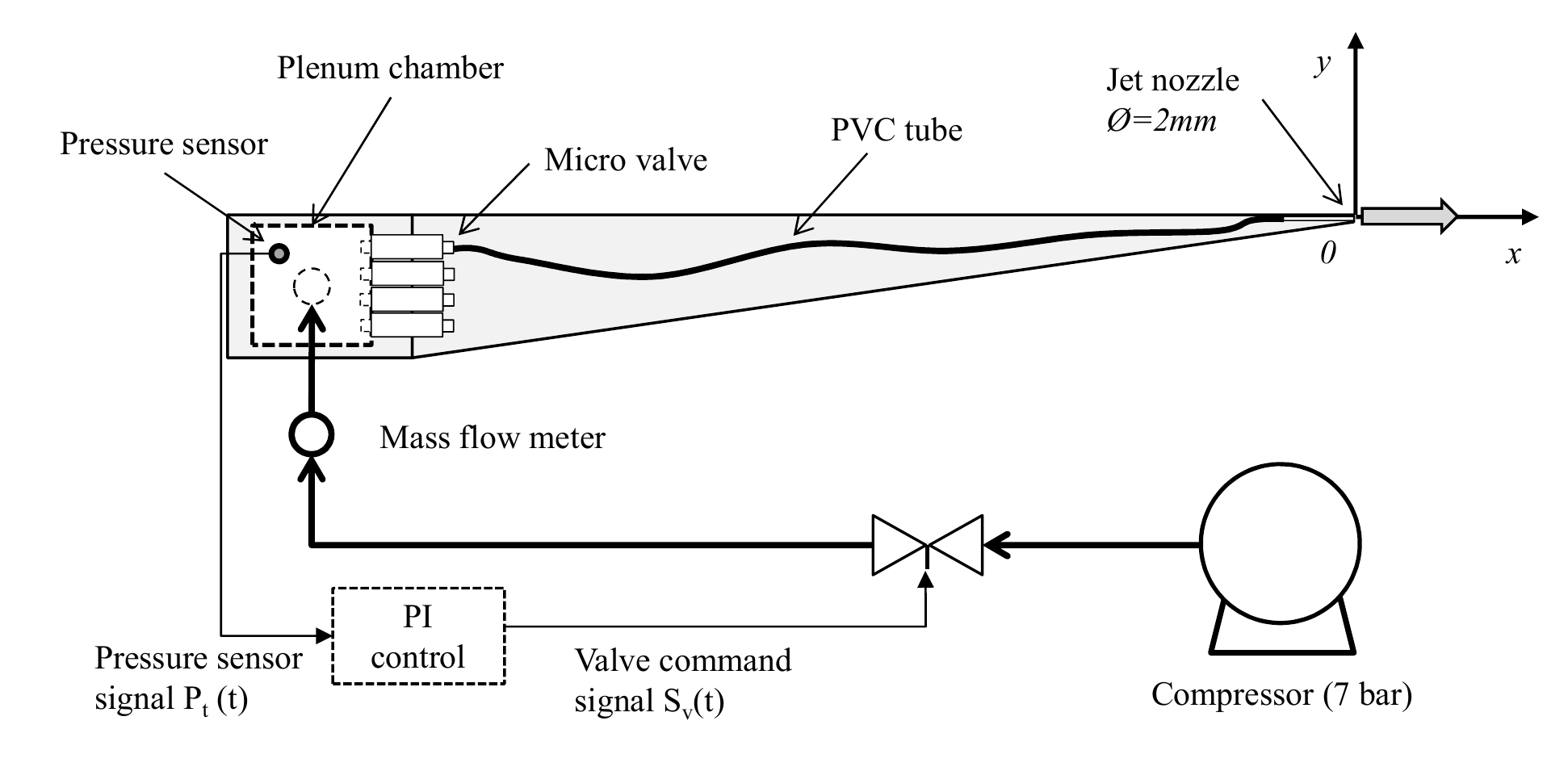}}% Images in 100% size
  \caption{Schematic representation of the splitter plate and the actuator system layout.}
\label{fig:actuator}
\end{figure*}

The actuator system comprises 96 micro jets at the trailing edge of the splitter plate, along its entire span. As shown in figure~\ref{fig:actuator} the jets actuate in the stream-wise direction at the origin of the mixing layer. The actuators function in a binary manner; they can be commanded to open and close at a maximum rate of \SI{500}{\hertz}. A plenum chamber, embedded in the splitter plate, directly supplies air to the micro-jets. The amplitude of actuation is globally adjusted by control of the pressure level $P_t$ in the plenum. For this purpose we use a PI controller which uses inputs from a differential pressure sensor, inside the plenum, and outputs command signals to an electrically controlled valve. The valve adjusts the flow rate of compressed air into the plenum chamber. The PI controller works at a rate of $\SI{100}{\hertz}$. 

Due to the long tubing connection between the control valve and the plenum, the system reaction time is on the order of seconds. Therefore, the actuation amplitude can only be maintained on a long time scale average and not instantaneously. It is typically used to maintain an average amplitude of actuation when switching the open-loop actuation frequency or duty cycle, or when switching between closed-loop control laws. We will discuss some effects of the different settings of gain of the actuation amplitude controller in section~\ref{sec:alt_pi}. 

An average amplitude of actuation can be represented in a non-dimensional form as the momentum coefficient:
\begin{equation}
C_{\mu}=\frac{\dot m_a}{\dot m_{\theta}},
\label{eq:c_mu}
\end{equation}

\noindent where $\dot m_a$ is the average mass flow rate of compressed air used by the actuator system. It is measured at the inlet of the plenum chamber by a {\it Brooks} 5863S mass flow meter. An estimation of the mean mass flow rate through the upper boundary layer is defined as $\dot m_{\theta}=\rho \, \theta \, l_z \, U_1$. Here, $\theta$ is the momentum thickness of the boundary layer, $\rho$ is the density of air, $l_z$ is the splitter plate span, and $U_1$ is the mean velocity of the upper stream. 

\begin{figure*}
  \centerline{\includegraphics[width=1\textwidth]{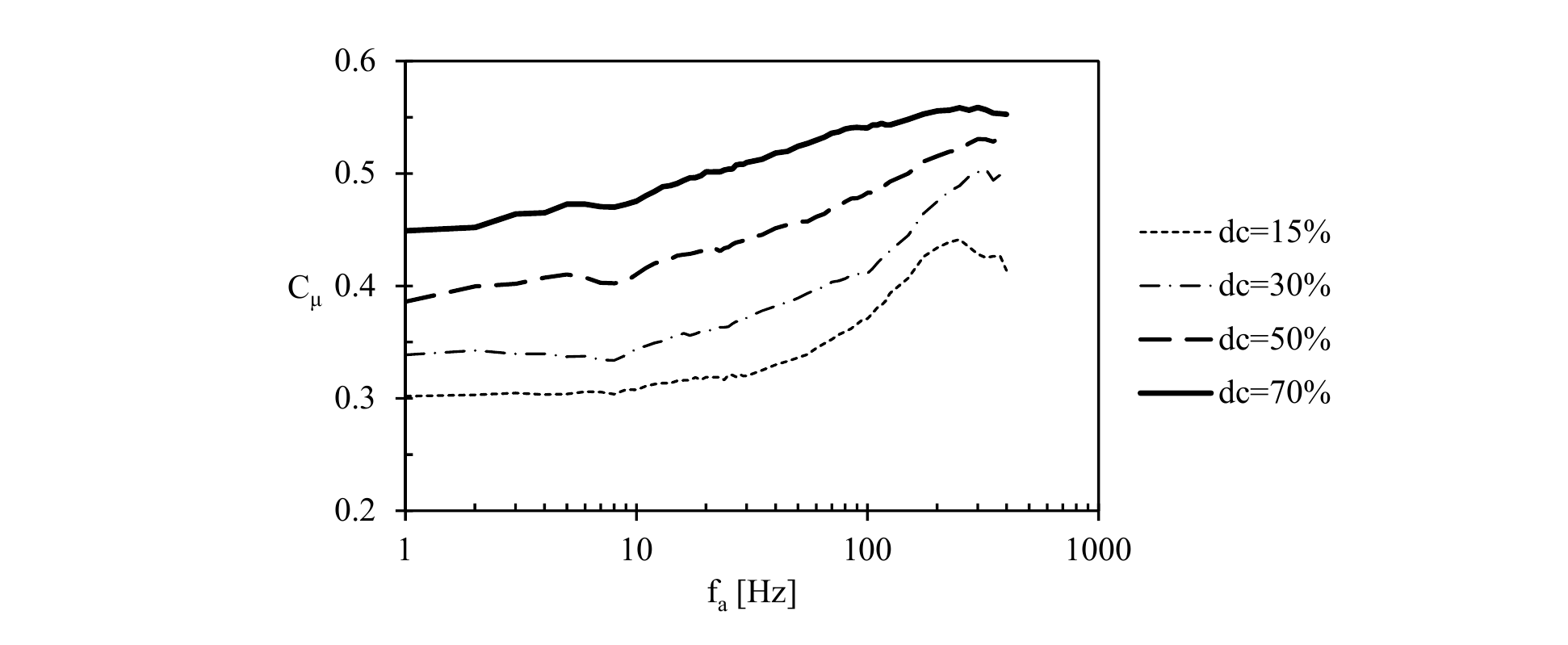}}% Images in 100% size
  \caption{Actuator system output $C_{\mu}$ for actuation frequencies $1<f_a<\SI{400}{\hertz}$, and duty cycles (dc) of $15\,\%$, $30\,\%$, $50\,\%$ and $70\,\%$, using a plenum pressure setting of $P_t=\SI{15}{\milli\bar}$. Actuation frequencies $f_a$ are plotted on a logarithmic scale, on the horizontal axis. Momentum coefficient $C_{\mu}$, plotted on the vertical axis, is computed for the low speed mixing layer conditions (see section~\ref{sec:natural}).}
\label{fig:cmu_dc}
\end{figure*}

The target amplitude of actuation is imposed by setting the plenum pressure to a desired level. Since the compressed air system cannot be controlled in-time, the instantaneous amplitude depends on the frequency of actuation ($f_a$) and the duty cycle (dc). Figure~\ref{fig:cmu_dc} shows the dependence of $C_{\mu}$, with respect to the available range of open-loop actuation settings. Based on these results we can approximate a constant average amplitude of actuation for $f_a \leq \SI{50}{\hertz}$, for a given constant duty cycle. These results are obtained for a plenum pressure level $P_t=\SI{15}{\milli\bar}$. This pressure level is used throughout this study except where stated otherwise. It corresponds to an average nozzle velocity of the jets of around \SI{3}{\meter\per\second}, which is very close to the convective speed of the mixing layer. Additional information on the velocity response of the jet actuators is available in~\cite{Parezanovic_FTC_2014}. 

\subsection{Sensor system}

The main state evaluation and control feed-back sensor system is a rake of up to 24 hot-wire probes. The hot-wire probes are of a classic single wire design, capable of measuring a modulus of velocity in a single plane ($xy$ plane in our case). In a mixing layer, the cross-stream component of velocity is one order of magnitude lower than the stream-wise component. Therefore, the contribution of the cross-stream component of velocity fluctuations $v^\prime$ can be neglected, and the hot-wire sensors can be considered sensitive only to the stream-wise component $u^\prime$. 

The rake can be positioned at key points of interest in the mixing layer flow by a $2D$ displacement console (in $xy$ plane) as illustrated in figure~\ref{fig:experiment}. The sensors cover a vertical length of up to \SI{184}{\milli\meter} depending on the number of wires used and their placement on the rake. Anemometer units are of in-house design using TSI 1750 constant temperature anemometry modules. They are calibrated for an optimal signal response up to $\SI{20}{\kilo\hertz}$. The hot-wire velocity measurements are sampled at a rate of $\SI{5}{\kilo\hertz}$, with a precision of $\SI{0.05}{\meter\per\second}$, and are corrected for temperature drifts using a reference temperature sensor at the inlet of the test section.

Two \textit{Lavision Image Intense} cameras coupled with a \textit{Spectra-Physics Lab-130} Nd:YAG laser are used for flow visualization in $xy$ plane. Seeding for visualization is provided by vaporized oil particles.

\subsection{Real-time I/O system}

Acquisition of data and real-time signal processing for control are performed by a \textit{Concurrent iHawk} Real-Time computing system. This system enables simultaneous data acquisition of up to 64 analog input channels and signal output to 96 digital output channels. The analog input channels are managed by two acquisition cards each with 32 channels. In the current experimental setup, the sensor signals are duplicated so that both cards acquire the same data from the hot-wire rake. One of the cards is dedicated to acquiring and relaying hot-wire signals as controller sensor input at a rate of \SI{1}{\kilo\hertz}, while the second card is recording the data for state evaluation and post-processing with a higher resolution of \SI{5}{\kilo\hertz}. The system uses in-house developed software for data recording, and Concurrent's Simulation Workbench software package for real-time I/O and processing operations.
% =============== CLOSED-LOOP CONTROL STRATEGY =======================%
\section{Closed-loop control strategy}\label{sec:strategy}

In this section we define the objective functions used for evaluation of both open- and closed-loop control. Next, we introduce Machine Learning Control (MLC) method and the basic principles of Genetic Programming (GP), on which this method is based. We then discuss specific aspects of implementation of MLC in the mixing layer experiment.  

\subsection{Objective functions}

The primary goal of the experiment is manipulation of the local flow properties of a turbulent mixing layer. The hot-wire sensors allow us to recover a local velocity fluctuation profile at the point of interest in the mixing layer. The total velocity $u_i$ measured by the $i$-th hot-wire is then decomposed in a sum of its mean and fluctuating parts, leading to: 
\begin{equation}
u^\prime_i(t)=u_i(t)-\langle u_i(t)\rangle_{\tau}
\label{eq:s}
\end{equation}
where
\begin{equation}
\langle u_i(t)\rangle_{\tau}=\frac{1}{\tau}\int_{t-\tau}^{t} u_i(t)\,\mathrm{d}t
\end{equation}
is the time-averaged mean velocity estimated on a period of $\tau=\SI{2}{\second}$, which corresponds to around 200 Kelvin-Helmholtz vortices.  

Based on these sensor inputs, we propose two different objective functions:
\begin{equation}
K=\sum_{i=1}^{N_s}
\left\langle
{u^\prime_i}^2(t)
\right\rangle_T
,
\label{eq:k}
\end{equation}
and
\begin{equation}
W=\frac{2K}
{
\displaystyle
\max_{i\in[1,N_s]}
\left\langle
{u^\prime_i}^2(t)
\right\rangle_T
}
.
\label{eq:w}
\end{equation}

\begin{figure*}
  \centerline{\includegraphics[width=1\textwidth]{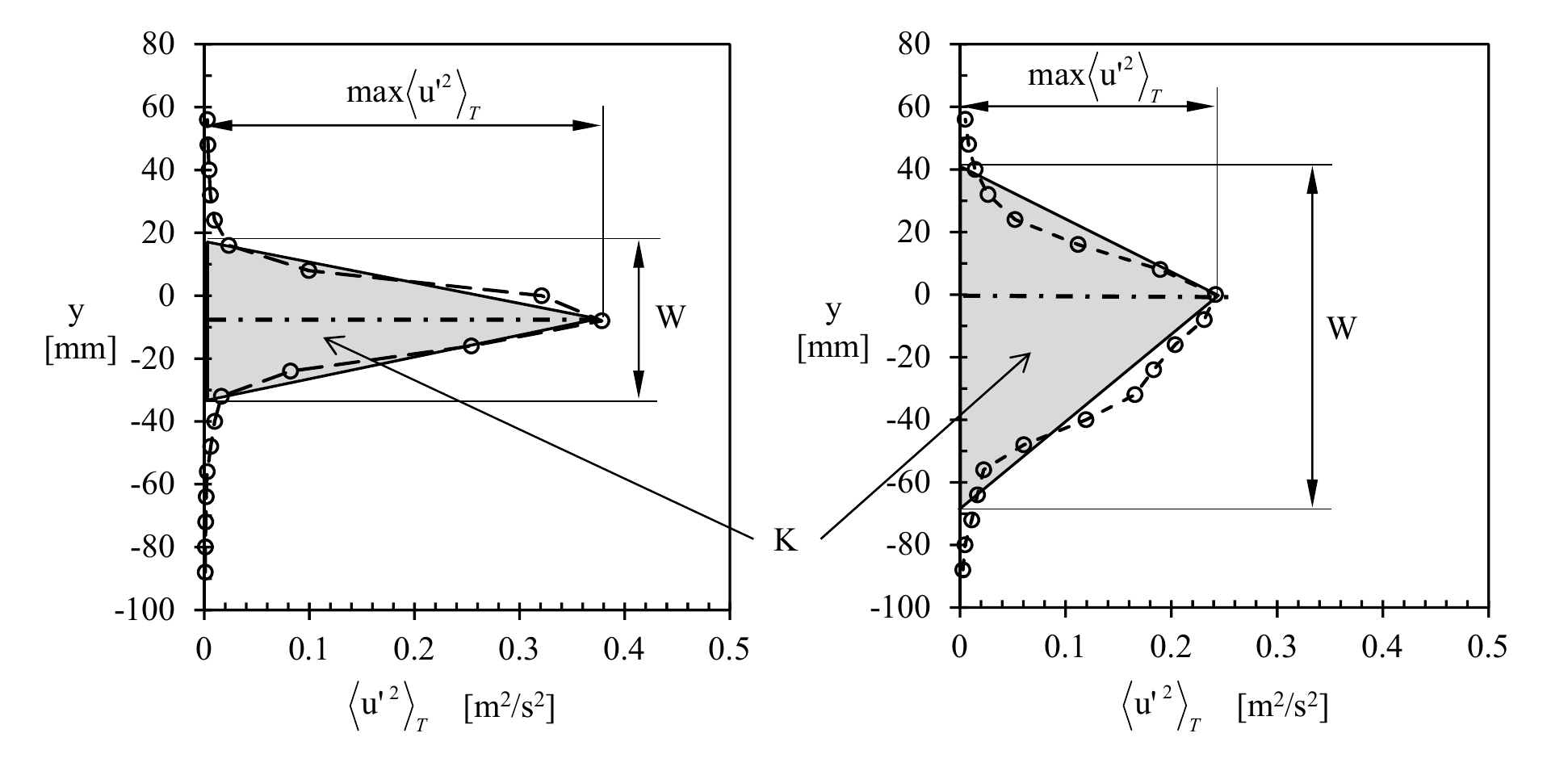}}% Images in 100% size
  \caption{Physical interpretation of the objective functions $K$ and $W$. Illustration on two different shapes of velocity variance profiles.}
\label{fig:objective}
\end{figure*}

In \eqref{eq:k} and \eqref{eq:w}, $N_s$ is the number of hot-wire probes used in the experiment (see section \ref{sec:exp_mlc}) for estimating the objective functions, and $T=\SI{10}{\second}$ is the averaging time for obtaining velocity variances $\langle{u^\prime_i}^2\rangle_T$. This value corresponds to a typical evaluation time of open-loop actuation or a closed-loop control law candidate. The objective function $K$ is calculated as the total sum of variances from all hot-wire sensors used. It is proportional to an integral $1D$ estimation of the turbulent kinetic energy in the mixing layer. The objective function $W$ is based on the local thickness (or width, hence $W$) of the mixing layer inferred from the velocity fluctuation profile shape. It is designed to favor small gradients in the distribution of fluctuation rates along the profile, while still seeking a large total sum of variances. This is brought on by assuming that the best mixing would be represented by a homogeneous fluctuation profile. $W$ is simply calculated by penalizing $K$ with the maximum value of variances detected in the profile. If $\displaystyle\max_{i}
\langle
{u^\prime_i}^2
\rangle_T$ 
is the height of a triangle which approximates the profile surface, then $W$ is proportional to the base of that triangle, or an effective local thickness of the mixing layer. This is illustrated in figure~\ref{fig:objective} for two examples of velocity variance profiles encountered in the mixing layer.  

For measuring the control effectiveness of the machine learning algorithm, we define non-dimensional cost functions based on $K$ and $W$ \textit{i.e.}  
\begin{eqnarray}
J_K=\frac{K}{K_u}\quad\text{and}\quad J_W=\frac{W}{W_u},
\label{eq:j}
\end{eqnarray}
where $K_u$ and $W_u$ are reference values of the objective functions for the un-actuated mixing layer. These cost functions correspond to a relative change, brought on by actuation, with respect to the un-actuated flow. Depending on the goal (see section~\ref{sec:results}), MLC seeks to maximize or minimize these cost functions, thereby enhancing or reducing the corresponding objective function, respectively.

These objective functions do not include any penalization terms regarding actuation cost. This choice was made taking into account the implementation of the PI control (see section~\ref{sec:actuators}) for maintaining an average constant actuation amplitude. In this case, instantaneous fluctuations of the actuation amplitude are acceptable, and can allow MLC to potentially evolve solutions qualitatively different from periodic or near-periodic forcing at a single frequency (see section~\ref{sec:alt_pi}). The energy efficiency of a control is evaluated, {\it a posteriori}, by considering the momentum coefficient $C_{\mu}$ given by~\eqref{eq:c_mu}.    

\subsection{Machine learning control}\label{sec:mlc}

We explore the possibilities of designing a closed-loop control strategy for strongly non-linear problems in fluid dynamics in a model-free manner. We are particularly interested in developing a sensor feedback approach where the control, to be determined, is a function of some sensor signals. This problem is equivalent to finding the regression model $f_\text{MLC}$ such that
\begin{equation}
b(t)= f_\text{MLC}\left(s_{i\in[1,N_s]}(t)\right)
\label{eq:b}
\end{equation}
where $b$ is the control law, and $s_i$ are the sensor signals. For determining $f_\text{MLC}$, we turn to the interdisciplinary field of machine learning \citep{Murphy_2012} and in particular to one of its well-established techniques for symbolic regression, called genetic programming \citep{koza1992genetic}. The advantage of symbolic regression over standard regression methods is that in symbolic regression, the search process works simultaneously on both the model specification problem and the problem of fitting coefficients. The genetic programming paradigm has been successfully employed to solve a large number of difficult problems such as pattern recognition, robotic control, theorem proving and air traffic control. To the authors' best knowledge, genetic programming has never been implemented in control of fluid flows by other authors.

\begin{figure}
  \centerline{\includegraphics[width=1\textwidth]{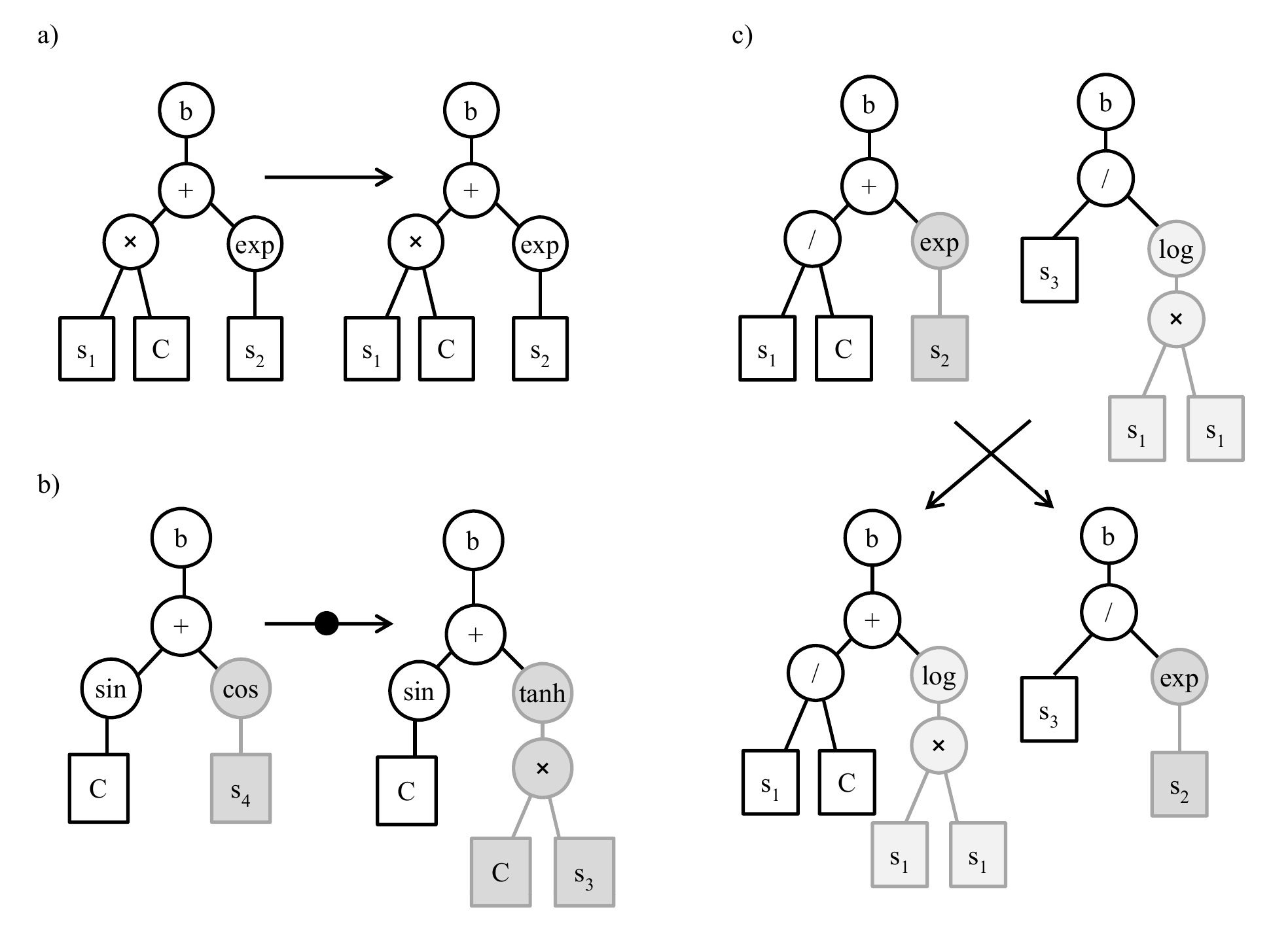}}% Images in 100% size
  \caption{Biologically-inspired operations performed by Genetic Programming (GP) on the tree-like
functions: a) replication, b) mutation and c) cross-over. The tree represented in a) corresponds to the function $b(t)=\text{C}\times s_1(t)+\exp\left(s_2(t)\right)$ where $\text{C}$ is a constant to be determined by GP. For the mutation b), the operation consists of selecting a node, erasing its subtree and replacing it by another one created randomly. Part of the information contained in the individual is kept while new information is allowed to enter the population. Mutation increases the diversity and is responsible for exploring the search space with large steps. For the cross-over c), one node in each of the two individuals selected for the cross-over is randomly chosen, the nodes and their subtrees are then exchanged. The cross-over is responsible for exploring the search space around well-performing individuals.}
\label{fig:genetic}
\end{figure}

Genetic programming (GP) is a biologically-inspired algorithm introduced by \citeauthor{koza1992genetic} to find computer programs that perform a user-defined task. It constitutes an extension of the genetic algorithm \citep{Goldberg_1989} for which the genetic operators (selection, cross-over and mutation) operate on a set of operations, elementary functions, variables and constants. When GP is used for symbolic regression, it combines automatically these elements to search for a symbolic expression that constitutes the best solution to a given optimization problem. Proposed symbolic expressions (examples shown in figure~\ref{fig:genetic}) are built as tree-like structures \citep{koza1992genetic} which can be easily evaluated in a recursive manner and described by a LISP expression. These trees are called individuals, while a population of individuals is a generation. GP is implemented as an iterative procedure in which a population of candidate solutions evolves toward better solutions by repeatedly undergoing genetic modifications based on their fitness.      

The first generation of individuals is created in a random manner. All individuals are evaluated and a fitness value is attributed based on how well they minimize or maximize the objective function. Based on these evaluations, a second generation of individuals is created using three principal genetic operations: replication, mutation and cross-over. Each of the genetic operations occurs with a pre-determined probability. If replication is selected then the candidate individual will simply be copied into the next generation (figure~\ref{fig:genetic}a). Mutation causes a part of the individual's tree to be replaced by a newly created random sub-tree as shown in figure~\ref{fig:genetic}(b). A cross-over operation involves a pair of individuals, which will exchange a randomly selected sub-tree between each other and two new individuals created in such a manner will become a part of the next generation (figure~\ref{fig:genetic}c). 

The selection process of candidate individuals uses the tournament method; few random individuals are chosen to compete in a tournament and the winner (based on its evaluated fitness) is selected for one of the genetic operations to be performed on it. The tournament is an efficient way of controlling the selectiveness of the GP evolutionary process. Usually, the size of the tournament is small compared to the size of the population. The bigger the tournament size $N_t$ is, the more probable it is to include the overall best individuals as participants, leading to a more selective process. The selection ranges from a random individual for $N_t=1$, to selection of the best individual of the generation if $N_t$ equals the size of the population of the current generation.
   
An additional operation available is elitism, which copies the single or few best individuals of one generation directly into the next generation, while avoiding the tournament process. This operation is generally used to ensure that the best individuals of one generation are not lost and remain available for further improvements in the future generations. 

When all the individuals of the next generation are created, they are evaluated and the selection process begins anew in order to build a subsequent generation. This iterative process continues for a desired number of generations. A rule of thumb is that given a sufficient number of individuals in each generation, a solution should be obtained in less than $N_g=50$ generations. There is no mathematical proof of convergence, but the method has been successful in many applications~\citep{lewis1992genetic,nordin1997line}.

\subsection{MLC in the mixing layer experiment}\label{sec:exp_mlc}

\begin{figure*}
  \centerline{\includegraphics[width=1\textwidth]{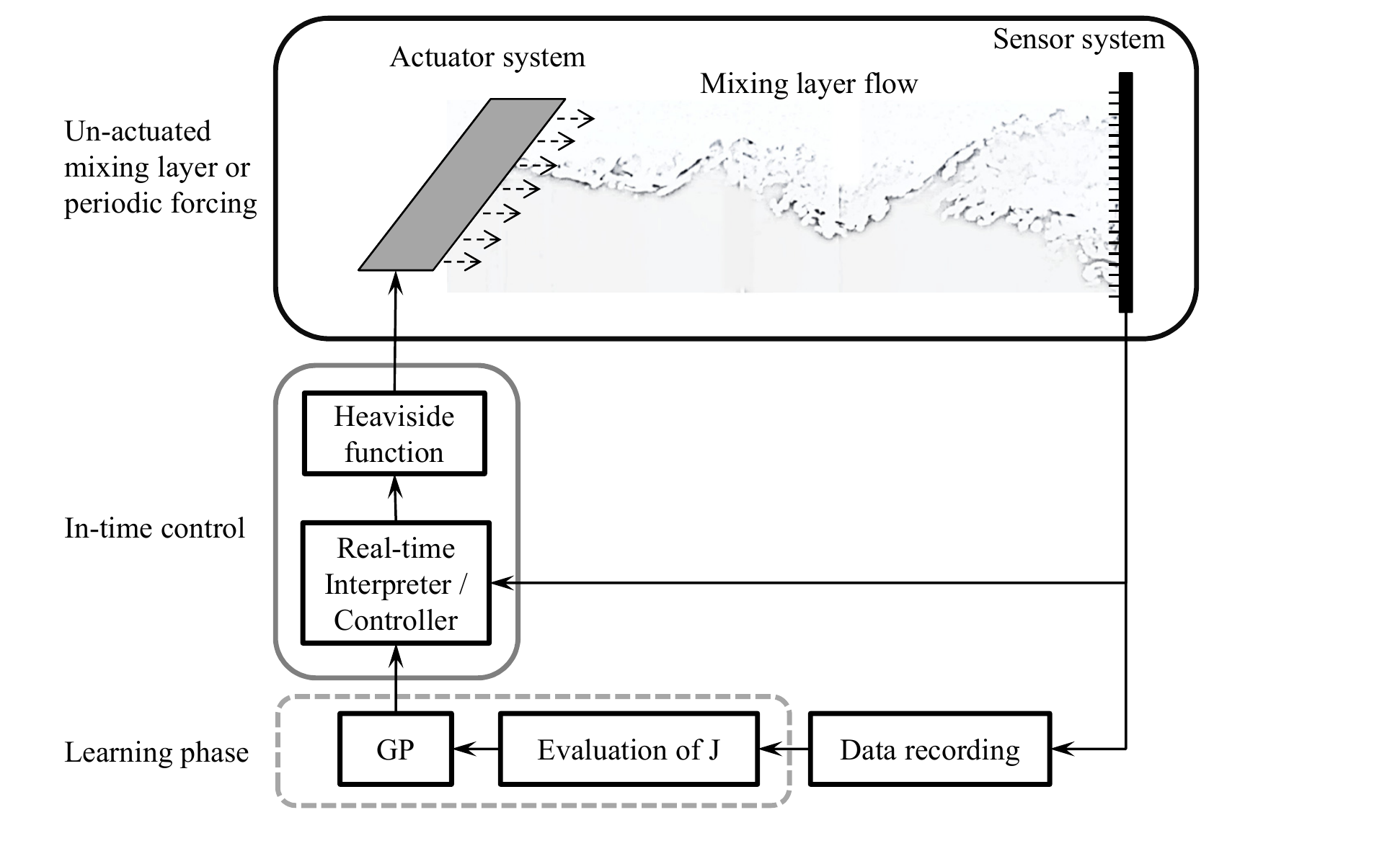}}% Images in 100% size
  \caption{The layout of MLC: the mixing layer plant with sensors and actuators, the in-time interpreter/controller and the MLC learning phase.}
\label{fig:controller}
\end{figure*}

The design layout of the experimental machine learning control is presented in figure~\ref{fig:controller}. We can arbitrarily separate the system into three parts for analysis. The top part is the experiment with the mixing layer plant and the associated actuator and sensor systems. This part represents the un-actuated flow and the elements sufficient for open-loop control; a function generator can be used to drive the micro-jets directly. 

The middle part is an interpreter/controller system which operates in real-time (cyclic rate of $\SI{1}{\kilo\hertz}$); sensor information is continuously acquired and evaluated through a control law to produce the control signal as in \eqref{eq:b}. An Heaviside function with a threshold at $0$ is applied to the evaluated signal in order to produce a binary command, which triggers the jet actuators. For minimum operation, this part requires an expression representing a control law (for example $b(t)=0$, to switch off actuation). This expression can be input manually, but the principal way of obtaining a control law expression is through MLC.     

The bottom part, shown in figure~\ref{fig:controller}, represents the learning phase. MLC sends a candidate individual to the real-time interpreter for testing. During the evaluation, data from the sensors are recorded simultaneously with data used for feedback. At the end of the evaluation time, a fitness value is calculated and assigned to the tested individual. This information is stored in the MLC database and the next candidate individual is sent for evaluation. The procedure is repeated until the final generation, from which the best individual is selected as the best closed-loop control law. This individual can then be applied continuously in the interpreter/controller and the learning phase is disconnected.

The genetic programming part of MLC code is developed in-house based on an open-source code \textit{ECJ} \citep{ecj}. Genetic programming codes using numerical simulations as evaluation data do not need to evaluate the same individual more than once. In the experiment, however, one is faced with uncertainties and stochasticity created by natural perturbations in the flow and the actuator system, measurement precision, etc. In order to be better adapted to experimental conditions, our MLC code contains crucial new features: 

\begin{enumerate}
%\begin{itemize}
  \item An individual is evaluated every time it appears in the genetic programming process.
  \item The cost of an individual is averaged with the values recorded previously for the same individual, resulting in a cumulative moving average being used as the cost function value.
  \item A predetermined number of best individuals in each generation are re-evaluated several more times to ensure a stable sample size for the averaging process.
  \item No duplicate individuals are allowed during the creation of the first generation.
%\end{itemize}
\end{enumerate}
\bigskip
The GP learning module thus assigns a cumulative average cost function value (fitness) to an individual. The objective is to choose as a best individual, one which has a continuously good performance, rather than one with strongly varying performance due to external perturbations. Such perturbations may cause an undeserved high fitness value to be attributed to an otherwise unremarkable individual, which is the only real danger to the convergence process. The averaged cost approach is designed to promote robustness and prevent such events from having an influence, but at a price of being much more conservative in the selection of individuals. If a duplicate individual is detected during the creation of the first generation, it is rejected and replaced with a new random individual. This ensures that the search space is not reduced by probing the same point two or more times.    

\begin{table}
  \begin{center}
\def~{\hphantom{0}}
  \begin{tabular}{lcc}
  	Parameter & Value \\[3pt]
    \hline
    Number of generations  & $N_g=$25 \\
    Population size & $N^{(1)}=1000$,  \\
					& $N^{(n)}=100$ ($n=2,\cdots,25$) \\
	Tournament size & $N_t=7$ \\				
    \hline
	Replication  & $P_r=0.1$ \\
	Cross-over  & $P_x=0.65$ \\
	Mutation & $P_m=0.25$ \\
	\hline
	Elitism & $N_e=1$ \\
	\hline
	Min. tree depth & 3 \\
	Max. tree depth & 10 \\
	\hline
	Operations & $+$, $-$, $\times$, $/$, $\sin$, $\cos$, $\exp$, $\tanh$, $\log$ \\ 
	
  \end{tabular}
  \caption{Standard set of MLC parameters used in the experiments (except where noted otherwise).}
  \label{tab:standard_params}
  \end{center}
\end{table}

Operations and elementary functions used in the experiment are: $+$, $-$, $\times$, $/$, $\sin$, $\cos$, $\exp$, $\tanh$ 
%$+,\,-,\,\times,\,/,\,\sin,\,\cos,\,\exp,\,\tanh$ 
and $\log$. The variables available to MLC are instantaneous values of velocity fluctuations from the hot-wire sensors \textit{i.e.} $s_i(t)\equiv u^\prime_i(t)$ in \eqref{eq:b}. Every third sensor in the rake is used as an MLC variable since more would only carry redundant information for the sensor-based controller. Hence, $N_s=21$ for the evaluation of the cost functions \eqref{eq:k} and \eqref{eq:w}, while only $7$ sensors across the shear layer are employed in \eqref{eq:b}. Finally, constants used are in a range of $[-1,1]$ with a precision up to a second decimal. The range and precision are adapted to be of the same order of magnitude as the typical velocity fluctuation information given by the sensors. 

In the experiment, the standard rates for the genetic programming operations are $10\,\%$ for replication, $25\,\%$ for mutation and $65\,\%$ for cross-over. Elitism is set to $N_e=1$, meaning a single best individual in a generation is sent directly to the next generation. An experiment consists of $N_g=25$ generations, where the first generation is composed of $N^{(1)}=1000$ individuals to allow for increased initial diversity, while subsequent generations ($n=2,\cdots,25$) contain $N^{(n)}=100$ individuals each. An individual is limited to a tree depth of $10$ levels. Standard MLC parameters are given in table~\ref{tab:standard_params}. Some variations of these settings are explored in section~\ref{sec:discussion}. 

The evaluation time of every individual in the experiment is $T=\SI{10}{\second}$, as mentioned earlier in this section. In addition, there is up to $\SI{6}{\second}$ of various delays in order to facilitate communication between the learning module, the controller and the data recording system. Also, a few seconds must be allowed for the mixing layer to register the effects of each new actuation when changing individuals during the learning process. Taking this into account, a single evaluation of an individual takes around $\SI{20}{\second}$ to complete. A standard MLC experiment (see table~\ref{tab:standard_params}) is completed in approximately $26$ hours. 

At the start of each generation, the un-actuated flow and the best open-loop actuation are re-tested in order to keep track of these reference values. Hence, each individual's cost ($J_K$ or $J_W$) is calculated using the reference baseline flow values ($K_u$ or $W_u$) obtained at the beginning of its own generation.

% =============== RESULTS ============================================%
 \section{Results}\label{sec:results}

In this section, we first present the initial conditions of the un-actuated mixing layer, for the Low-Speed (LS) and High-Speed (HS) configurations. This is followed by a detailed mapping of the impact of open-loop forcing on the cost functions $J_K$ and $J_W$. We will then focus on the best closed-loop control laws found by MLC for several different cases. Maximization of cost functions $J_K$ (MaxK) and $J_W$ (MaxW) is explored at $x=\SI{200}{\milli\meter}$ and $x=\SI{500}{\milli\meter}$, as well as minimization of $J_K$ (MinK) at $x=\SI{200}{\milli\meter}$. The impact of different flow conditions is explored with a maximization experiment of $J_W$ for the HS mixing layer configuration, at $x=\SI{200}{\milli\meter}$. Lastly, the effect of alternative gain settings of the actuation amplitude controller is tested in an MLC experiment. These settings cause the amplitude controller to become affected by actuation, and MLC is capable of exploiting this coupling to produce a different type of a solution. In sections dealing with MLC results, values for the un-actuated flow, best open-loop reference and the best individual are evaluations performed in the final generation of each experiment. Therefore, they represent results of single evaluations of the mentioned configurations. Statistically converged results of these cases will be discussed in section~\ref{sec:discussion}.

\subsection{Un-actuated flow}\label{sec:natural}

Two configurations of the mixing layer are used in the experiment: a low-speed (LS) mixing layer with stream velocities of $U_1=\SI{4.7}{\meter\per\second}$ and $U_2=\SI{1.3}{\meter\per\second}$, and a high-speed (HS) configuration with stream velocities of $U_1=\SI{9}{\meter\per\second}$ and $U_2=\SI{1.7}{\meter\per\second}$. In each configuration, the magnitude of velocity $U_1$ is selected to have a stable boundary layer on the upper surface of the splitter plate, laminar in the LS configuration, and turbulent for the HS setting. 

The two streams of the LS configuration produce a mixing layer with a velocity ratio of $r=0.27$, and a convective velocity of $U_c=\SI{3}{\meter\per\second}$. The boundary layer on the high speed side of the splitter plate is laminar ($\delta_{99} = \SI{6.9}{\milli\meter}$, $\theta = \SI{0.73}{\milli\meter}$ and $H = 2.44$). Based on the momentum thickness of the upper boundary layer, the initial Reynolds number of the mixing layer can be estimated as $\Rey_{\theta}=500$. 

The HS configuration features a turbulent upper boundary layer ($\delta_{99} = \SI{13.3}{\milli\meter}$, $\theta = \SI{0.81}{\milli\meter}$ and $H = 1.55$). Initial conditions for this configuration are $\Rey_{\theta}=2000$, $U_c=\SI{5.4}{\meter\per\second}$ and $r=0.18$. 

\begin{figure}
  \centerline{\includegraphics[width=1\textwidth]{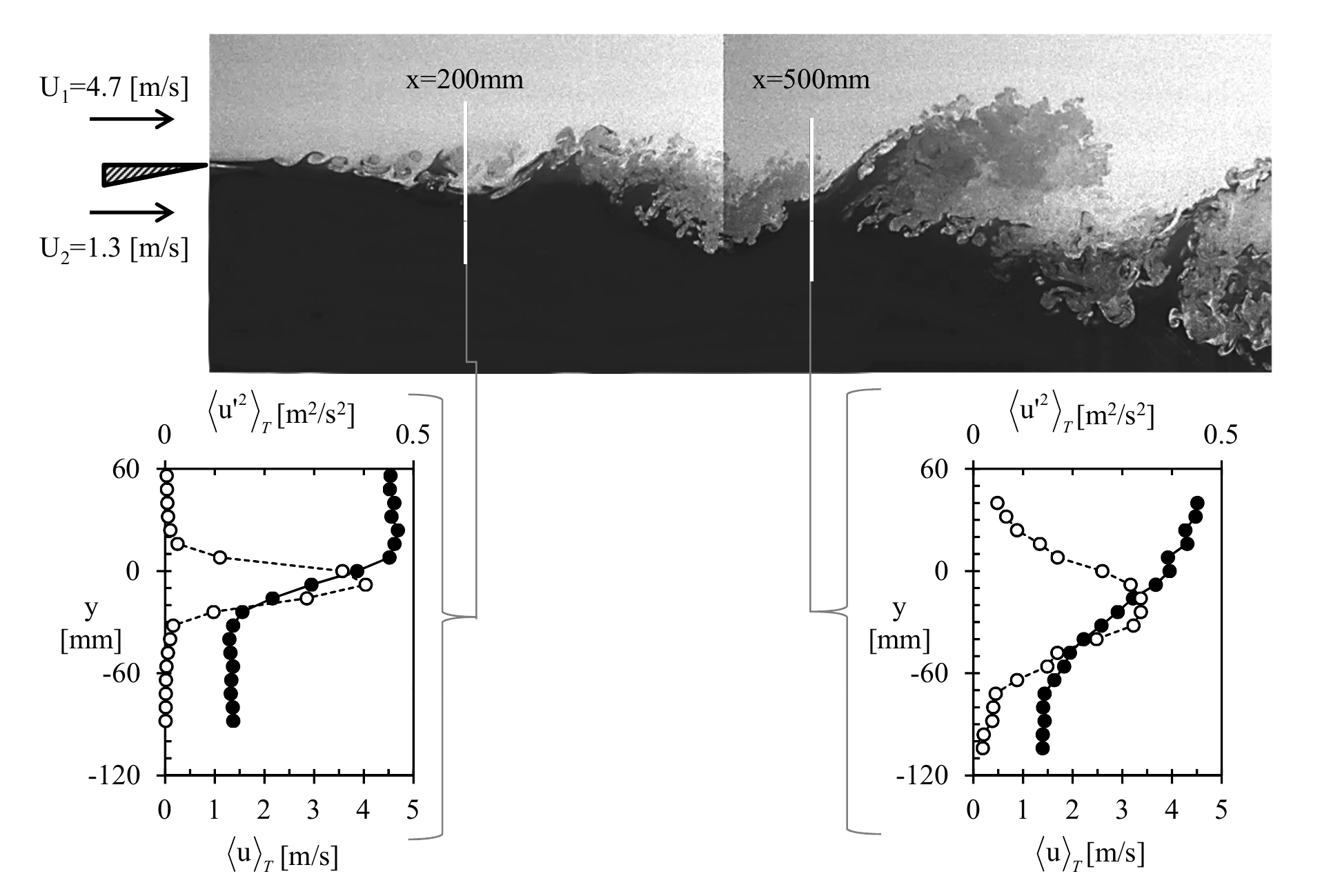}}% Images in 100% size
  \caption{Smoke visualization of the un-actuated low-speed mixing layer with mean velocity $\langle u\rangle_T$ and velocity variance $\langle
{u^\prime}^2
\rangle_T$ profiles for sensor positions at $x=\SI{200}{\milli\meter}$ and $x=\SI{500}{\milli\meter}$.}
\label{fig:smoke_visu}
\end{figure}

In both mixing layer configurations, the thickness of the boundary layer on the lower side of the splitter plate is $\delta_{99} < \SI{1}{\milli\meter}$. This is a result of very low velocity $U_2$ (in both cases) and the inserted foam flow stabilizer. The impact of this boundary layer on the development of the mixing layer can be considered negligible. The main focus in this section will be on the low-speed configuration. The high-speed configuration will be examined in section~\ref{sec:max_w_hv}, and later discussed with respect to control robustness in section~\ref{sec:robust}. 

Based on the flow visualization of the low-speed mixing layer, shown in figure~\ref{fig:smoke_visu}, we can estimate the frequency of the initial Kelvin-Helmholtz instability as $f_{\text{KH}}=\SI{90}{\hertz}$. Two streamwise locations are shown in figure~\ref{fig:smoke_visu}, one at $x=\SI{200}{\milli\meter}$ and the second at $x=\SI{500}{\milli\meter}$, with their respective mean velocity and velocity variance profiles. These locations are used throughout this study and are selected as points of interest based on mixing layer spatial evolution data presented in more detail in \citet{Parezanovic_FTC_2014}.

\subsection{Periodic forcing}\label{sec:periodic}

The response of the mixing layer to periodic forcing is evaluated using objective functions $J_K$ and $J_W$ on the entire frequency range $1<f_a<\SI{400}{\hertz}$ available to the actuator system. Measurements are performed for the sensor system placement at $x=\SI{200}{\milli\meter}$ and $x=\SI{500}{\milli\meter}$. Actuation frequency is changed in varying steps (roughly a logarithmic distribution) using $60$ points in the $\SI{400}{\hertz}$ range. Four different actuation duty cycles are used as mapping parameters: $15\,\%$, $30\,\%$, $50\,\%$ and $70\,\%$.

\begin{figure*}
  \centerline{\includegraphics[width=1\textwidth]{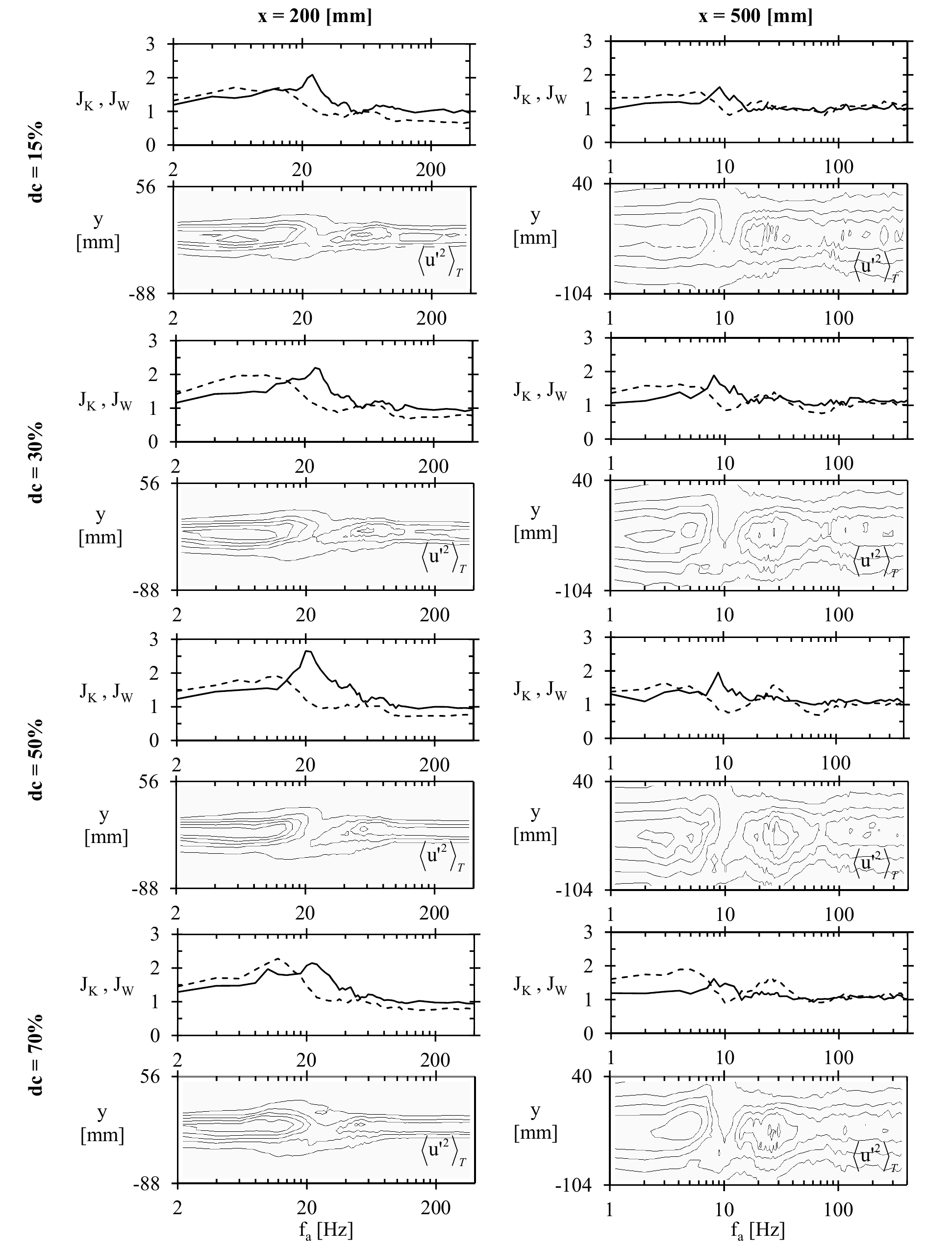}}% Images in 100% size
  \caption{Mapping of mixing layer response to open-loop actuation for $f_a=0-\SI{400}{\hertz}$ at $x=\SI{200}{\milli\meter}$ (left) and $x=\SI{500}{\milli\meter}$ (right). Duty cycle settings (top to bottom): $dc=15\,\%$, $dc=30\,\%$, $dc=50\,\%$ and $dc=70\,\%$. All diagrams share a common horizontal axis denoting $f_a$ on a logarithmic scale. Cost functions are represented with continuous line for $J_W$ and dashed line for $J_K$. Corresponding mean velocity variance profiles for each case are plotted as iso-lines of $\langle{u^\prime}^2\rangle_T$, with five levels in range $[0.1\,0.5]\,(\si{m^{2}.s^{-2}})$. Vertical axis for profile plots corresponds to the vertical position of the sensors.}
\label{fig:j_vs_f_200_500}
\end{figure*}

Figure~\ref{fig:j_vs_f_200_500} shows the evolution of the cost functions $J_K$ and $J_W$ with regard to open-loop actuation frequency when sensors are placed at $x=\SI{200}{\milli\meter}$ and $x=\SI{500}{\milli\meter}$. The values for the un-actuated case are equal to $1$. For the sensor location at $x=\SI{200}{\milli\meter}$ (left in figure~\ref{fig:j_vs_f_200_500}), the most striking increase of $J_W$ can be observed for $dc_{\text{\tiny Wmax}}^{\text{\tiny OL}}=50\,\%$ in a very narrow band of actuation frequencies around $f_{\text{\tiny Wmax}}^{\text{\tiny OL}}=\SI{21}{\hertz}$. This result corresponds to a thick velocity variance profile in the associated diagram. Values of $J_K$ are increased the most for $f_{\text{\tiny Kmax}}^{\text{\tiny OL}}=\SI{12}{\hertz}$ and $dc_{\text{\tiny Kmax}}^{\text{\tiny OL}}=70\,\%$. For higher actuation frequencies of $f_a>\SI{75}{\hertz}$ the values of $J_K$ are reduced below the value of the un-actuated flow in all four cases. This result corresponds to high frequency forcing effect on the near-field of a free shear layer, reported by~\cite{vukasinovic2010}. However, a relation $J_W>1$ stands for the whole range of tested actuation frequency and duty cycle combinations.

Actuation effects measured further downstream at $x=\SI{500}{\milli\meter}$ are similarly shown in figure~\ref{fig:j_vs_f_200_500}(right). A highest gain with regard to $J_W$ is again obtained for $dc_{\text{\tiny Wmax}}^{\text{\tiny OL}}=50\,\%$, but this time it is at a lower frequency $f_{\text{\tiny Wmax}}^{\text{\tiny OL}}=\SI{9}{\hertz}$. Evolution of $J_K$ indicates two positive extremum points, with the highest gain on the lower frequency extremum with the actuation parameters $f_{\text{\tiny Kmax}}^{\text{\tiny OL}}=\SI{5}{\hertz}$, $dc_{\text{\tiny Kmax}}^{\text{\tiny OL}}=70\,\%$. The minimum for $J_K$ is obtained around $f_{\text{\tiny Kmin}}^{\text{\tiny OL}}=\SI{75}{\hertz}$ for $dc_{\text{\tiny Kmin}}^{\text{\tiny OL}}=30\,\%$ and $50\,\%$, although it is not as pronounced as for the measurements at $x=\SI{200}{\milli\meter}$. $W$ is always larger than the un-actuated flow value in this case as well. These optimal forcing parameters will be used in the MLC experiments as best actuation settings $f_a^\text{opt}$ and $dc^\text{opt}$ for reference open-loop control values, and are given in table~\ref{tab:res}. 

\subsection{MLC maximization of the width (MaxW experiment)}\label{sec:max_w}

MLC control is first applied to search for a maximization of the objective function $J_W$ at $x=\SI{200}{\milli\meter}$ and $x=\SI{500}{\milli\meter}$. These optimization problems are denoted MaxW. The GP parameters are identical in both cases and correspond to standard parameters given in table~\ref{tab:standard_params}. The best performing individuals for the two experiments are selected in the last MLC generation \textit{i.e.} $n=25$. 

% They have the following expressions:
%
% $(sin (sin (- (exp (- (* S21 (exp (* S3 S3))) S18)) (cos (* (exp S15) (* (* (* S3 S3) S21) S18))))))$ (200)
%
% $(* S21 (log (- (cos (- (sin S21) (log (+ (sin S21) S21)))) (log (+ (sin S21) (sin S21))))))$ (500)
%

% \begin{figure}
%     \begin{center}
%         \begin{tabular}{lcc}
% 			\rotatebox{90}{\phantom{000000000}$x=\SI{200}{\milli\meter}$} & 
%             \begin{overpic}[width=1\textwidth]{figures/200_wmax_res}
%             \end{overpic}\\
%             & \\
%             \rotatebox{90}{\phantom{000000000}$x=\SI{500}{\milli\meter}$} &
%             \begin{overpic}[width=1\textwidth]{figures/500_wmax_res}
%             \end{overpic}\\    
%         \end{tabular}
%         %\vspace{-.05in}
%         \caption{Results of MaxW experiments at $x=\SI{200}{\milli\meter}$ and $x=\SI{500}{\milli\meter}$: a) profiles of $\langle{u^\prime}^2(t)\rangle_T$ for the un-actuated flow (dashed line), open-loop reference (white circles) and MLC (black circles), b) actuation signal spectra for open-loop (grey) and MLC (black) and c) pseudo-visualization using hot-wire time series of $u^\prime$ (gray-scale) in range $[-1.5,1.5]$ (\si{\meter\per\second}) with a stripe depicting actuation activity (black stands for actuation).}
%         \label{fig:200_wmax_res_500_wmax_res}
%     \end{center}
% \end{figure}

\begin{figure}
\subfigure[$x=\SI{200}{\milli\meter}$.\label{fig:200_wmax_res}]{
\centerline{\includegraphics[width=0.95\textwidth]{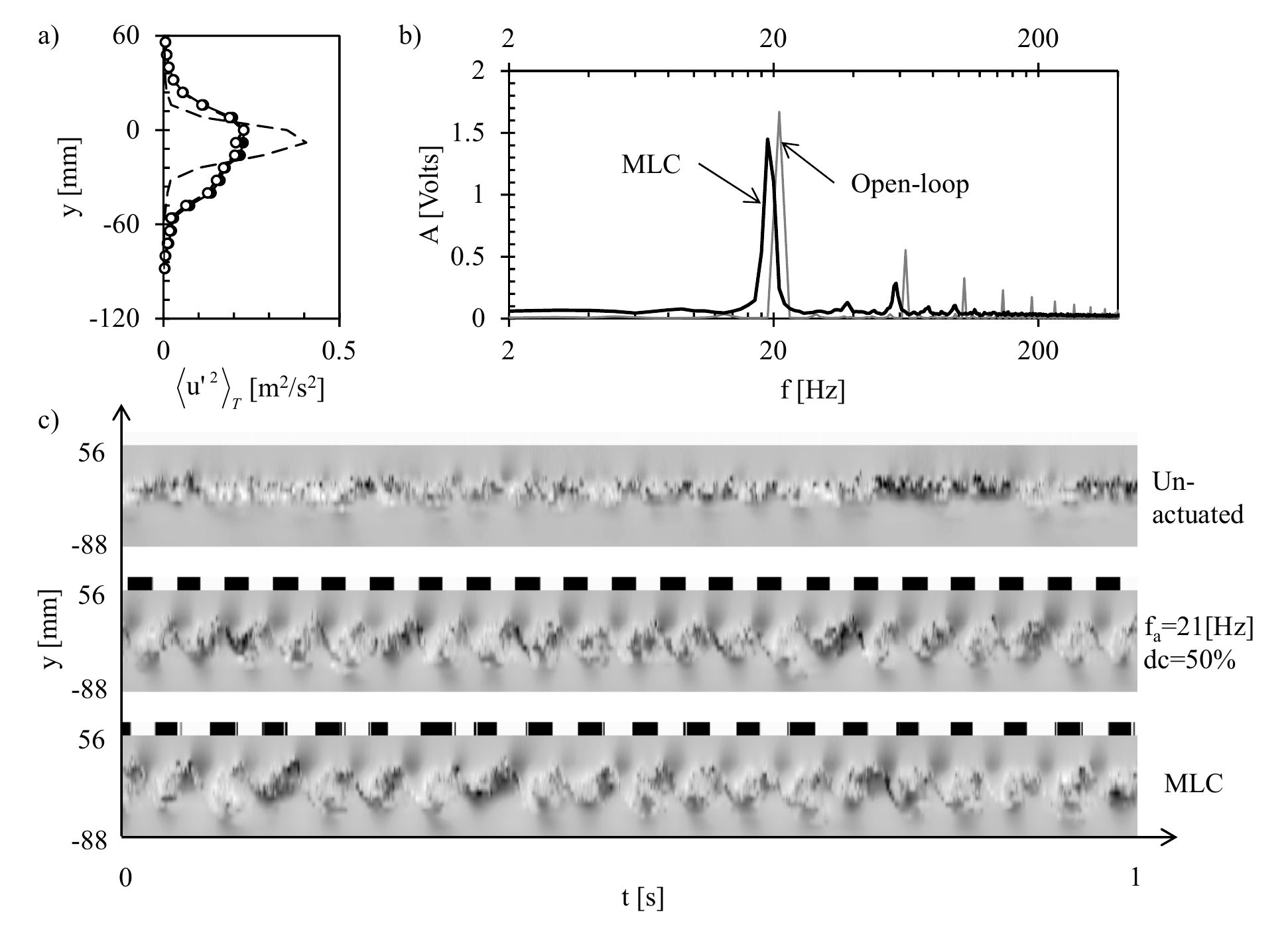}}% Images in 100% size
}
\subfigure[$x=\SI{500}{\milli\meter}$.\label{fig:500_wmax_res}]{
\centerline{\includegraphics[width=0.95\textwidth]{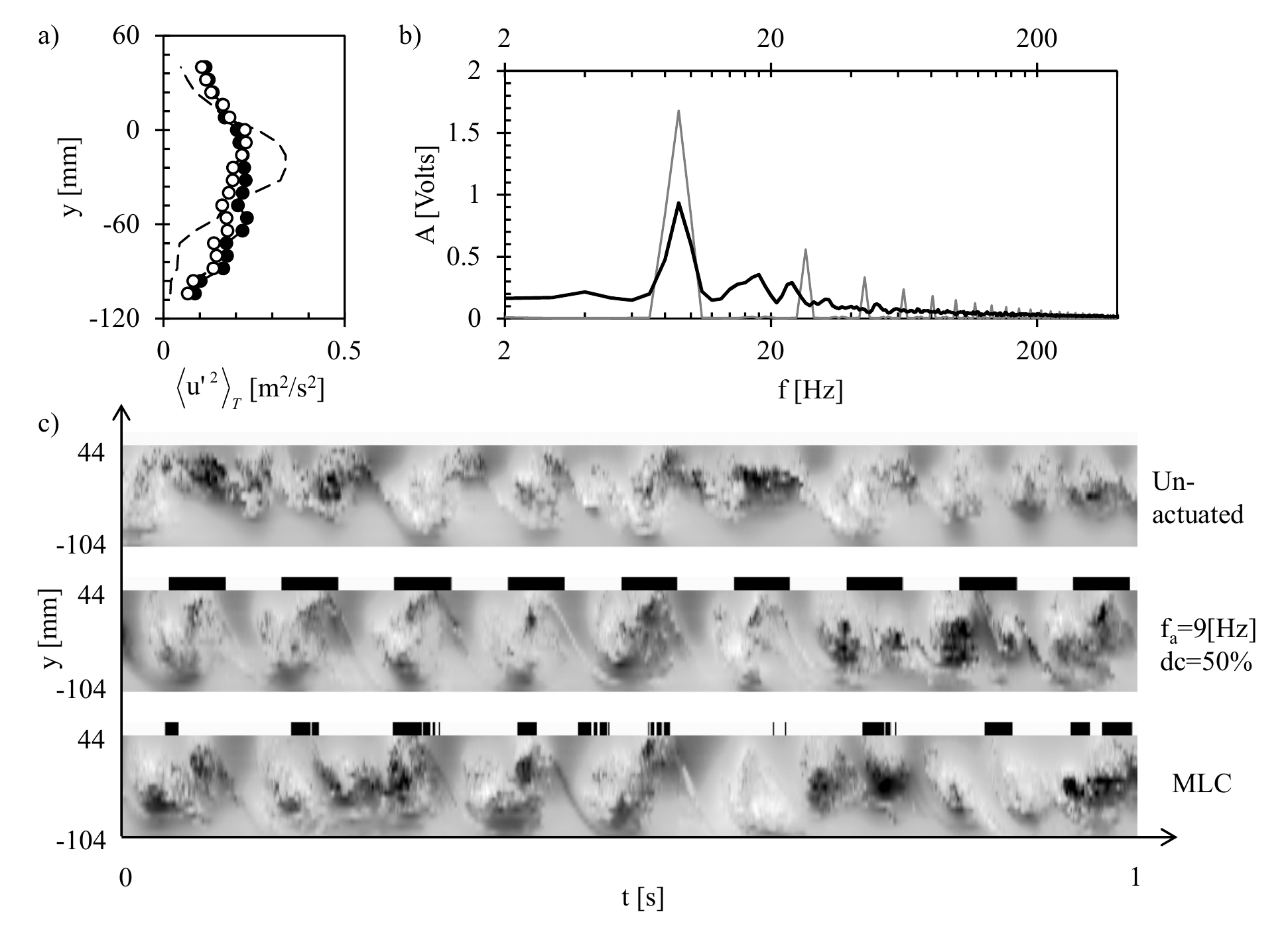}}% Images in 100% size
}
\caption{Results of MaxW experiments: a) profiles of $\langle{u^\prime}^2(t)\rangle_T$ for the un-actuated flow (dashed line), open-loop reference (white circles) and MLC (black circles), b) actuation signal spectra for open-loop (grey) and MLC (black) and c) pseudo-visualization using hot-wire time series of $u^\prime$ (gray-scale) in range $[-1.5,1.5]$ (\si{\meter\per\second}) with a stripe depicting actuation activity (black stands for actuation).}
\label{fig:200_wmax_res_500_wmax_res}
\end{figure}

Figure~\ref{fig:200_wmax_res}(a) shows velocity variance profiles for the un-actuated flow, best open-loop actuation and the best MLC individual. We can note that the fluctuation profiles of the two actuated cases are almost indistinguishable from each other but significantly different from the un-actuated flow profile. Figure~\ref{fig:200_wmax_res}(b) shows the FFT spectra of the actuation signals for the open- and closed-loop control. This allows us to determine a dominant actuation frequency $f_{\text{\tiny Wmax}}^{\text{\tiny MLC}}$ for the best closed-loop individual and compare with the best (fixed) open-loop actuation frequency $f_{\text{\tiny Wmax}}^{\text{\tiny OL}}$. It can be noted that in this experiment the best closed-loop had a dominant frequency $f_{\text{\tiny Wmax}}^{\text{\tiny MLC}}=\SI{19}{\hertz}$ slightly lower than the best open-loop frequency $f_{\text{\tiny Wmax}}^{\text{\tiny OL}}=\SI{21}{\hertz}$. Figure~\ref{fig:200_wmax_res}(c) shows pseudo-visualization of the respective flows, based on Taylor's hypothesis of frozen turbulence, using the velocity fluctuation $u^\prime$ time-series from hot-wire probes. The time series represented corresponds to $\SI{1}{\second}$ which is equivalent to a spatial length of around $10$ times of the sensor rake height used in this experiment. It can be concluded that the best MLC individual performs very similarly to the best open-loop control. They both increase the local thickness of the mixing layer, as evidenced from the velocity variance profiles. The visualization reveals very similar sizes and frequencies of coherent structures.

Results are displayed in a similar fashion in figure~\ref{fig:500_wmax_res} for the $x=\SI{500}{\milli\meter}$ case. This time, there are some differences in the velocity variance profiles (figure~\ref{fig:500_wmax_res}(a)) between the controlled cases, but it can be said that both work to create a similar mixing layer thickness increase. Spectra in figure~\ref{fig:500_wmax_res}(b) show the MLC actuation frequency is identical to open-loop setting: $f_{\text{\tiny Wmax}}^{\text{\tiny MLC}}=f_{\text{\tiny Wmax}}^{\text{\tiny OL}}=\SI{9}{\hertz}$. However, from the actuation stripe on top of pseudo-visualization fields, we can observe a much different instantaneous duty cycle  for the closed-loop case. 

Table~\ref{tab:res} provides quantitative information to compare the optimal parameters corresponding to MaxW for the two measurement locations. Actuation frequency and duty cycle are fixed quantities in the case of open-loop actuation. For the closed-loop case, the reported actuation frequency is the dominant frequency in the spectrum of the best closed-loop actuation signal \textit{i.e.} $f_{\text{\tiny Wmax}}^{\text{\tiny MLC}}$. The duty cycle reported for this case is estimated as the percentage of total evaluation time for which the actuators are in the active state. The momentum coefficient $C_{\mu}$ represents the invested energy of actuation. Finally, the values for the cost function $J_W$ are given.

It can be concluded that the best closed-loop individuals can outperform their open-loop counterparts in all aspects. In particular, an interesting case is the optimal duty cycle for closed-loop control at $x=\SI{500}{\milli\meter}$ which is selected around $25\,\%$. This corresponds well in comparison with the open-loop response maps in section~\ref{sec:periodic}, where figure~\ref{fig:j_vs_f_200_500} shows that the maximum of $J_W$ at $x=\SI{500}{\milli\meter}$ is similar for $dc=30\,\%$ and $dc=50\,\%$. Duty cycle $dc=25\,\%$ was not tested in open-loop mapping experiments, but the best MLC individual points in the direction of that result. However things may not be that simple since in the case of $x=\SI{200}{\milli\meter}$, MLC has found repeatedly that the best resulting closed-loop actuation has a dominant frequency of $\SI{19}{\hertz}$ with an optimal duty cycle of $48\,\%$. This duty cycle is very close to the optimal open-loop value of $50\,\%$, yet the frequency for open-loop should not be optimal below at least $\SI{20}{\hertz}$.  

\subsection{MLC maximization of the fluctuation energy (MaxK experiment)}\label{sec:max_k}

Results for the maximization of the energy-based objective function $K$ are presented here for $x=\SI{200}{\milli\meter}$ and $x=\SI{500}{\milli\meter}$. Resulting best individuals are selected from generation $n=25$ from standard MLC experimental runs.

\begin{figure}
\subfigure[$x=\SI{200}{\milli\meter}$.\label{fig:200_kmax_res}]{
\centerline{\includegraphics[width=0.95\textwidth]{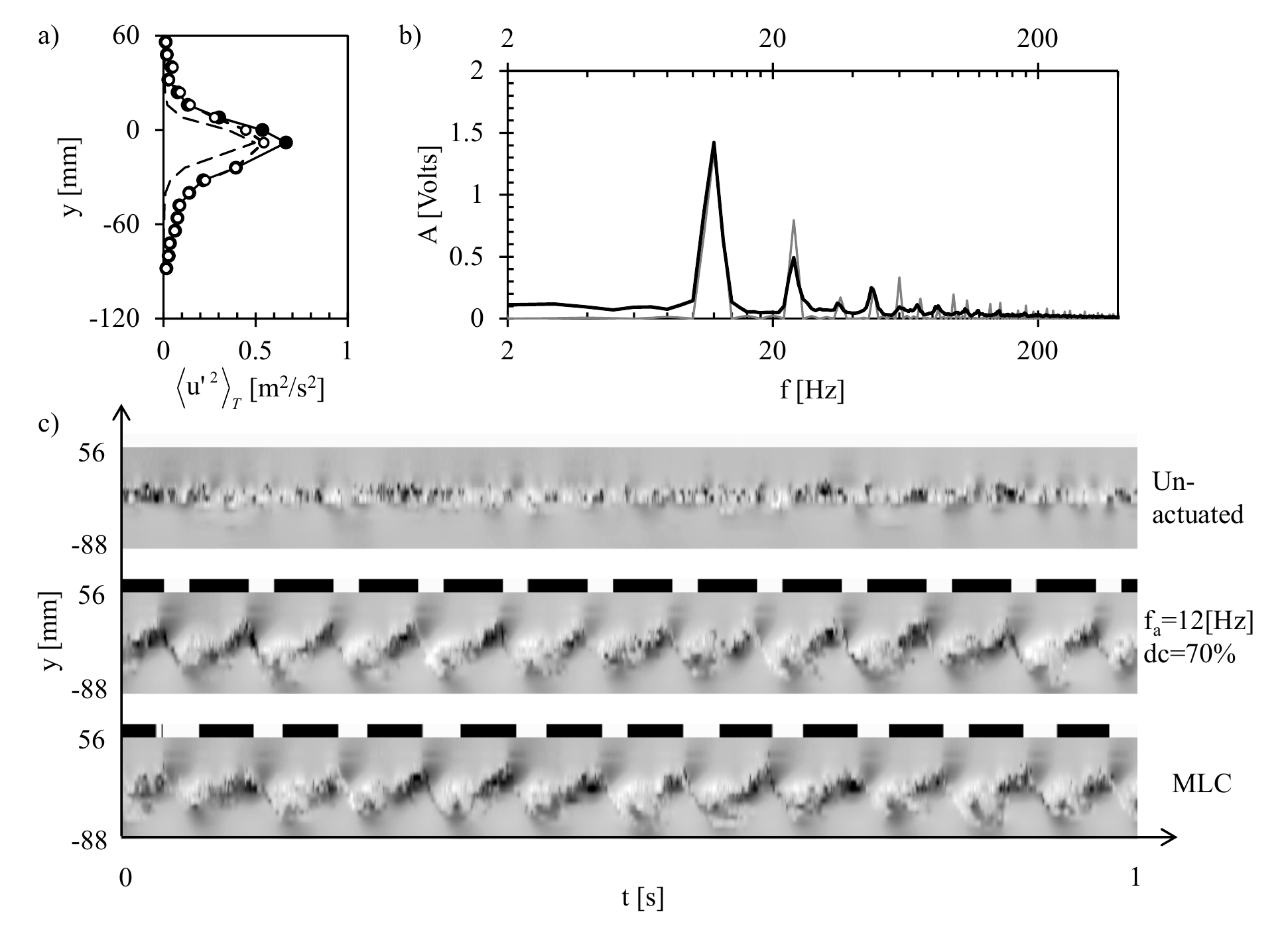}}% Images in 100% size
}
\subfigure[$x=\SI{500}{\milli\meter}$.\label{fig:500_kmax_res}]{
\centerline{\includegraphics[width=0.95\textwidth]{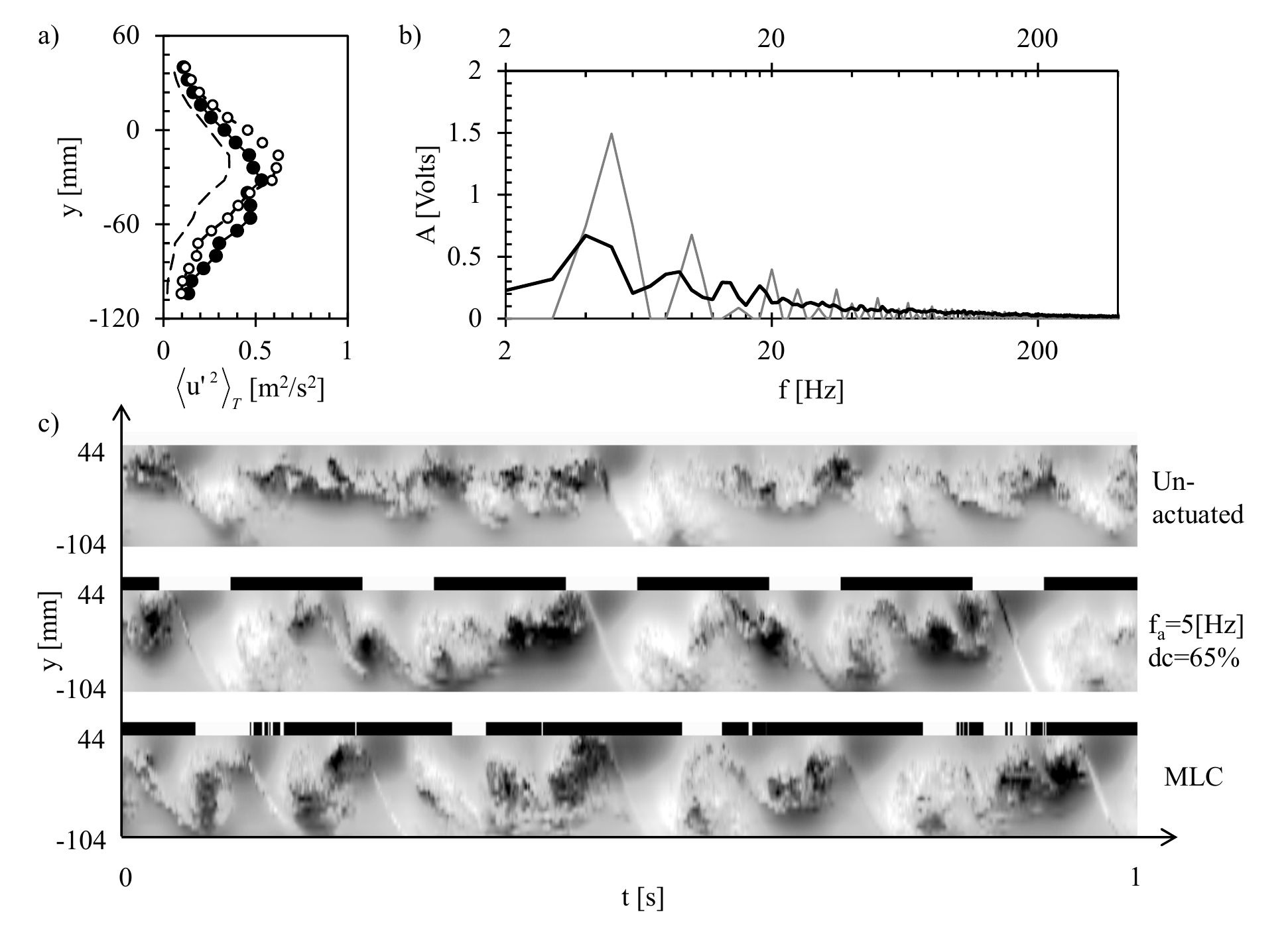}}% Images in 100% size
}
\caption{Results of MaxK experiments (see caption of figure~\ref{fig:200_wmax_res_500_wmax_res} for details).}
\label{fig:200_kmax_res_500_kmax_res}
\end{figure}

Mean velocity variance profiles shown in figure~\ref{fig:200_kmax_res}(a) for the results obtained at $x=\SI{200}{\milli\meter}$, reveal a slight advantage for the closed-loop control, while the global frequency for closed-loop control matches the open-loop value, shown in figure~\ref{fig:200_kmax_res}(b). Visualization fields in figure~\ref{fig:200_kmax_res}(c) indicate no significant differences between the two controls, at first glance. Data in table~\ref{tab:res} under the $x=\SI{200}{\milli\meter}$ entry for MaxK show a slight optimization of the closed-loop duty cycle which could be responsible for improved performance.

Results are similarly presented in figure~\ref{fig:500_kmax_res} for maximization of $K$ at $x=\SI{500}{\milli\meter}$. Here (figure~\ref{fig:500_kmax_res}a) we note larger differences in the velocity variance profile shapes between the controlled cases. Actuation frequency is slightly lower for closed-loop (shown in figure~\ref{fig:500_kmax_res}b). Pseudo-visualization in figure~\ref{fig:500_kmax_res}(c) reveals a large increase of the actuation duty cycle for the case of the closed-loop control. Results for MaxK at $x=\SI{500}{\milli\meter}$, in table~\ref{tab:res}, show that open-loop is slightly better and much more efficient in terms of invested energy.      

\begin{table}
  \begin{center}
\def~{\hphantom{0}}
  \begin{tabular}{lcccc}
  & \multicolumn{2}{c}{Maximization of $W$ (MaxW)}& & \\
  & & & &  \\
      $x=200\,[\si{\milli\meter}]$ & $f_a^\text{opt}\,[\si{\hertz}]$ & $dc^\text{opt}\,[\%]$ & $C_{\mu}$ & $J_W$ \\[3pt]
       Open-loop (OL) & 21 & 50 & 0.25 & 2.13 \\
       Closed-loop (MLC) & 19 & 48 & 0.25 & 2.25 \\
       &  & & & \\
      $x=500\,[\si{\milli\meter}]$ & $f_a^\text{opt}\,[\si{\hertz}]$ & $dc^\text{opt}\,[\%]$ & $C_{\mu}$ & $J_W$ \\[3pt]
       Open-loop (OL) & 9 & 50 & 0.20 & 1.54 \\
       Closed-loop (MLC) & 9 & 25 & 0.13 & 1.70 \\   
  & & &  & \\
  & \multicolumn{2}{c}{Maximization of $K$ (MaxK)}& & \\  
  & & & & \\
       $x=200\,[\si{\milli\meter}]$ & $f_a^\text{opt}\,[\si{\hertz}]$ & $dc^\text{opt}\,[\%]$ & $C_{\mu}$ & $J_K$ \\[3pt]
       Open-loop (OL) & 12 & 70& 0.27 & 2.27 \\
       Closed-loop (MLC) & 12 & 62 & 0.24 & 2.44 \\
       & & & & \\
      $x=500\,[\si{\milli\meter}]$ & $f_a^\text{opt}\,[\si{\hertz}]$ & $dc^\text{opt}\,[\%]$ & $C_{\mu}$ & $J_K$ \\[3pt]
       Open-loop (OL) & 5 & 70 & 0.25 & 1.98 \\
       Closed-loop (MLC) & 4.5 & 80 & 0.34 & 1.95 \\  
  \end{tabular}
  \caption{Results of the final generation of the MLC maximization experiments of $W$ (MaxW) and $K$ (MaxK) for two positions of the sensors $x=\SI{200}{\milli\meter}$ and $x=\SI{500}{\milli\meter}$. The optimal values of $f_a^\text{opt}$ and $dc^\text{opt}$ used for the open-loop reference were determined in section \ref{sec:periodic}.}
  \label{tab:res}
  \end{center}
\end{table}

\subsection{MLC minimization of the fluctuation energy (MinK experiment)}\label{sec:min_k}

\begin{figure}
\subfigure[Minimization of $K$ (low-speed).\label{fig:200_kmin_res}]{
\centerline{\includegraphics[width=0.95\textwidth]{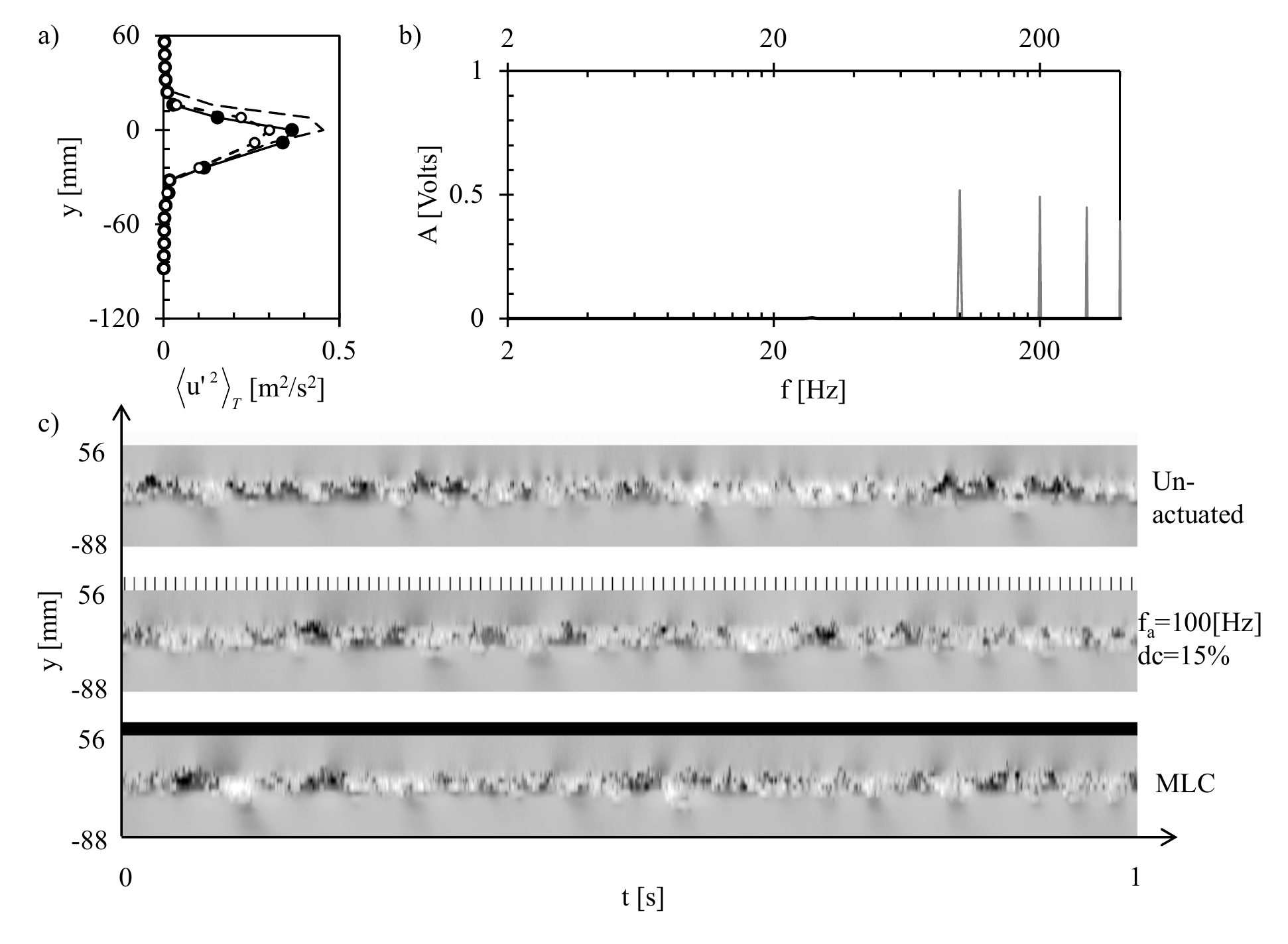}}% Images in 100% size
}
\subfigure[Maximization of $W$ (high-speed).\label{fig:200_wmax_res_hv}]{
\centerline{\includegraphics[width=0.95\textwidth]{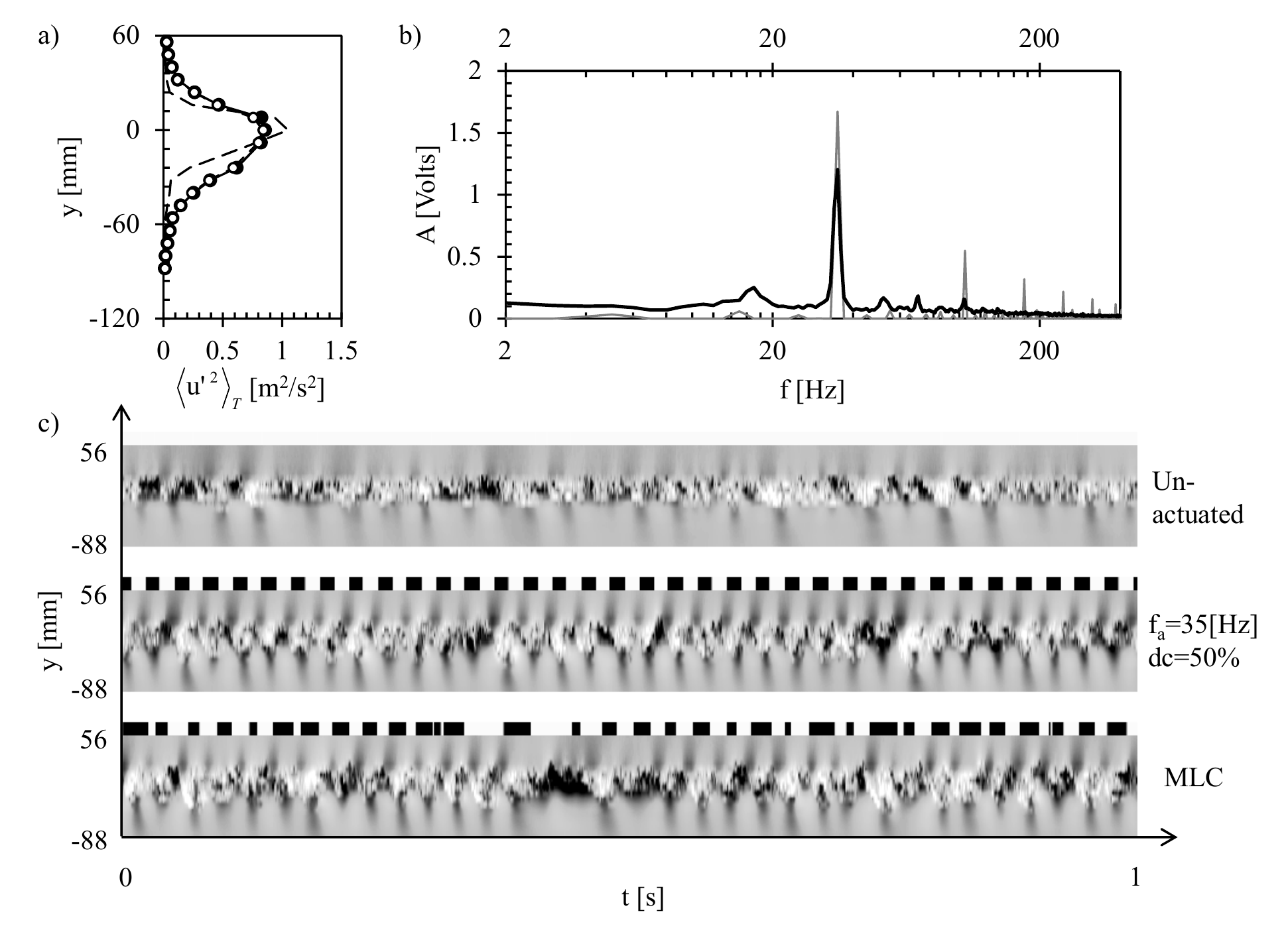}}% Images in 100% size
}
\caption{Results of MinK (low-speed) and MaxW (high-speed) experiments at $x=\SI{200}{\milli\meter}$ (see caption of figure~\ref{fig:200_wmax_res_500_wmax_res} for details).}
\label{fig:200_kmin_res_200_wmax_res_hv}
\end{figure}

In section~\ref{sec:periodic} it has been shown that the total energy of fluctuations in the mixing layer could be reduced using high frequency periodic actuation. An MLC experimental run is performed in search of the best $K$ minimization closed-loop control and results are shown in figure~\ref{fig:200_kmin_res}. The velocity variance profiles in figure~\ref{fig:200_kmin_res}(a) show that both control approaches have a much smaller relative impact (compared to maximization experiments) when minimization of the mixing layer energy is attempted. The actuation frequency of the best MLC individual cannot be detected in figure~\ref{fig:200_kmin_res}(b). Figure~\ref{fig:200_kmin_res}(c) reveals that continuous actuation was selected as the best closed-loop control. Resulting average reductions of $K$ are $31\,\%$ for open-loop and $25\,\%$ for closed-loop control.

An MLC experiment seeking minimization of $K$ has also been performed at $x=\SI{500}{\milli\meter}$, but is not shown here. In this case, continuous actuation or no actuation are alternatively found as best individuals. This results from the fact that both have very similar objective function values at $x=\SI{500}{\milli\meter}$, \textit{i.e.} continuous actuation has some effect closer to the trailing edge but further downstream this effect is dispersed and the mixing layer instability re-appears. Note that high-frequency open-loop actuation still has the ability to reduce $K$ at $x=\SI{500}{\milli\meter}$ as shown in section~\ref{sec:periodic}.

\subsection{MLC maximization of the width in a high-speed mixing layer}\label{sec:max_w_hv}

The maximization of $W$ has been explored in high-speed mixing layer flow conditions. The high-speed mixing layer is fully turbulent from its inception, however both low- and high-speed configurations are turbulent in nature at the sensor location at $x=\SI{200}{\milli\meter}$ downstream of the trailing edge. Still, the local frequency and velocity fluctuation levels are significantly different compared to the low-speed case. The mapping experiment of open-loop response for the high-speed configuration is performed and yields the best actuation parameters as $f_{\text{\tiny Wmax}}^{\text{\tiny OL}}(\text{HS})=\SI{35}{\hertz}$ and $dc_{\text{\tiny Wmax}}^{\text{\tiny OL}}(\text{HS})=50\,\%$, which are used as reference. Complete mapping results are omitted here for brevity.  

Resulting velocity variance profiles and actuation frequencies, shown in figure~\ref{fig:200_wmax_res_hv}(a) and (b), are very similar for both control types. Visualization in figure~\ref{fig:200_wmax_res_hv}(c) reveals that closed-loop control is less regular that in previously documented cases for this measurement location, in low-speed conditions. This attests to the more turbulent nature of the high-speed mixing layer. The best MLC individual shows performance very close to best open-loop: $J_{\text{\tiny Wmax}}^{\text{\tiny OL}}(\text{HS})=1.72$ and $J_{\text{\tiny Wmax}}^{\text{\tiny MLC}}(\text{HS})=1.74$.

\subsection{Influence of the high-gain amplitude controller on the MLC}\label{sec:alt_pi}

Some properties of the PI controller, used for actuation amplitude control, were discussed in section~\ref{sec:actuators}. In the early experiments, a higher gain was used for the PI controller, making it very sensitive. Combined with the slow response of the compressed air distribution system, the effect was such that a low-frequency modulation introduced into the actuation signal caused the actuation amplitude to be significantly increased past the desired level on a short time scale, while preserving the long time scale target average amplitude.

\begin{figure*}
  \centerline{\includegraphics[width=1\textwidth]{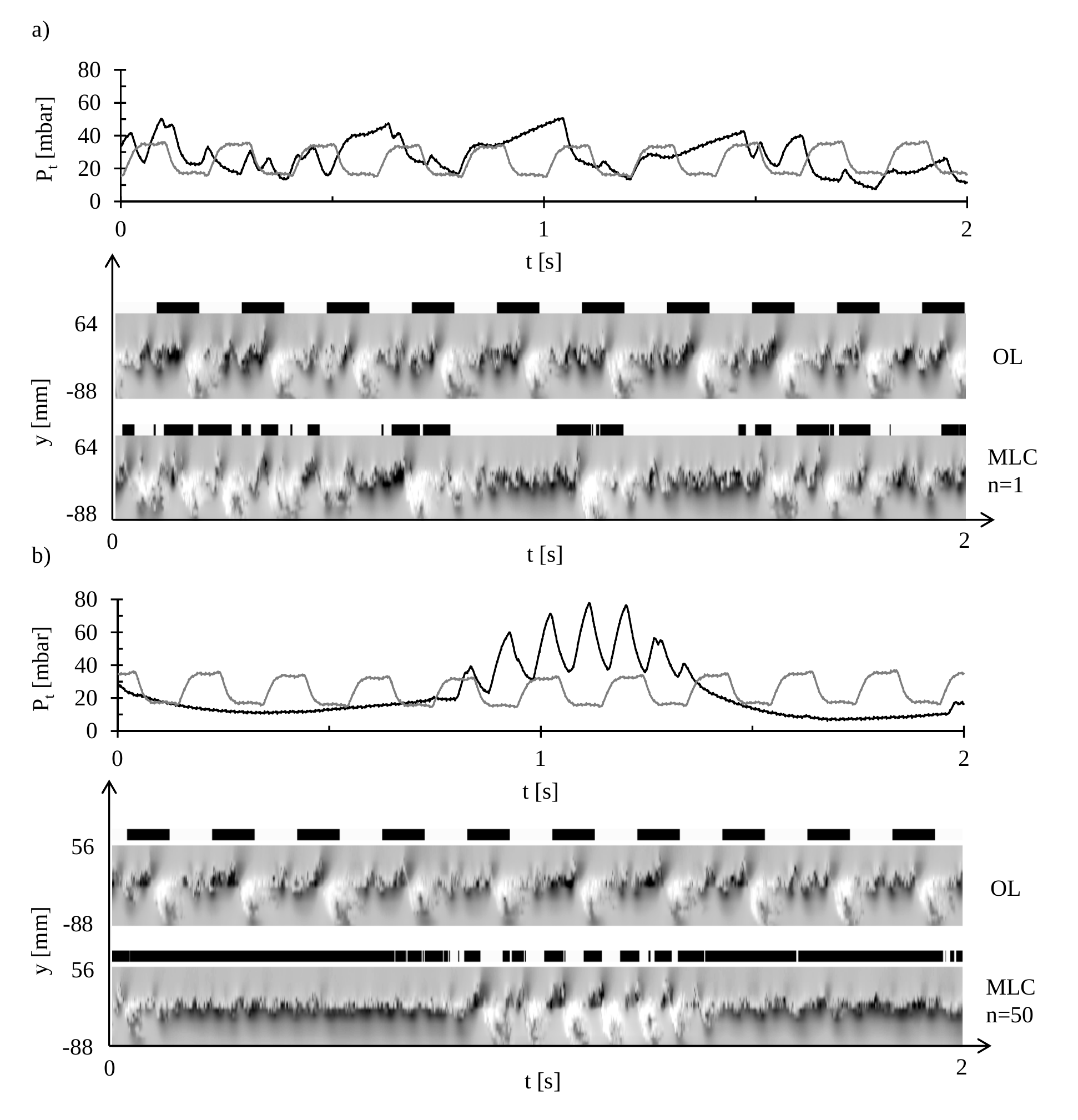}}% Images in 100% size
  \caption{Results of MLC experiment with the high-gain amplitude controller for the open-loop reference and best MLC individual in generation a) $n=1$ and b) $n=50$. Top diagrams show instantaneous pressure in the plenum of the actuator system, for the open-loop case (grey line) and best MLC individual (black line). Bottom diagrams show corresponding pseudo-visualizations and actuation activity (black and white stripes). Greyscale map of visualization fields corresponds to $u^\prime$ in range $[-1.5,1.5]$ (\si{\meter\per\second}).}
\label{fig:pt_ecj_visu}
\end{figure*}

Figure~\ref{fig:pt_ecj_visu} shows two results from this experiment: (a) is the best individual from the first generation and (b) the best individual from generation $n=50$. In each case the upper plot shows the instantaneous pressure levels in the plenum chamber of the actuator system, which corresponds to the pseudo-visualization time series shown in the lower plots. Except for the number of generations $N_g$, other parameters used in this experiment correspond to standard values in table~\ref{tab:standard_params}.

The first generation individual (figure~\ref{fig:pt_ecj_visu}a) yields a similar value of $J_K$ as the best open-loop actuation. However, due to the sensitive PI control, there exists a better solution (figure~\ref{fig:pt_ecj_visu}b), which employs bursts of actuation followed by comparatively long periods of continuous blowing. The action of continuous blowing drains the plenum chamber and the PI controller increases the flow rate to compensate. When the actuation burst starts, \textit{i.e.} at the first short closure of the actuators, an over-pressure is created in the plenum. Each cycle in the burst is then injecting much more momentum into the mixing layer, causing a high instantaneous actuation amplitude, which yields a larger value of $K$. As the PI controller catches up with this process, another period of continuous blowing starts and the cycle repeats indefinitely.

This sequence comprises an actuation frequency of around $\SI{12}{\hertz}$ (corresponding to the "burst mode"), modulated by a very low frequency $f_{mod}<\SI{1}{\hertz}$ (corresponding to the continuous blowing periods). The resulting cost function $J_K$ values are $J_{\text{\tiny Kmax}}^{\text{\tiny OL}}=3.04$ and $J_{\text{\tiny Kmax}}^{\text{\tiny MLC}}=3.93$. In the case of this experiment a higher average amplitude of actuation was set for both control approaches ($P_t=\SI{25}{\milli\bar}$), leading to momentum coefficient values of $C_{\mu}^{\text{\tiny OL}}=0.58$ and $C_{\mu}^{\text{\tiny MLC}}=0.89$. More information on the GP iterative process, leading to this solution, are given in section~\ref{sec:gp_params}.

% =============== DISCUSSION OF MLC PERFORMANCE ======================%
\section{Discussion of MLC performance}\label{sec:discussion}

% \comment{In this section, I only modified the caption of the figure and some other notations.}

In this section we discuss the convergence and capabilities of the MLC process in finding the best control law candidates. We explore the sensitivity of MLC to different parameters, such as: the size of the first and subsequent generations and settings for genetic operations. The statistically converged results are considered in the comparison between the performance of the best individuals and the reference open-loop actuations. Robustness to different flow conditions, for both open- and closed-loop control, is discussed. Finally, we examine the costs of MLC and other methods in terms of experimental time needed.

%\vspace*{0.5cm}
%\hrule
%\vspace*{0.5cm}

\subsection{Influence of the variation of GP parameters}\label{sec:gp_params}

Two crucial questions naturally arise when dealing with any stochastic optimization method: i) what is the best obtainable solution and do we have any guarantee to obtain it, and ii) what are the optimal parameters for obtaining this solution? Due to the nature of the GP process, it is impossible to respond to the first question when a completely unknown system is concerned. In our case, however, we have at least a partial response in the form of the reference open-loop actuation. Parameters of GP we selected to vary were the population size and the balance of probabilities assigned to each of the three genetic operations.

\begin{figure}
  \centerline{\includegraphics[width=1\textwidth]{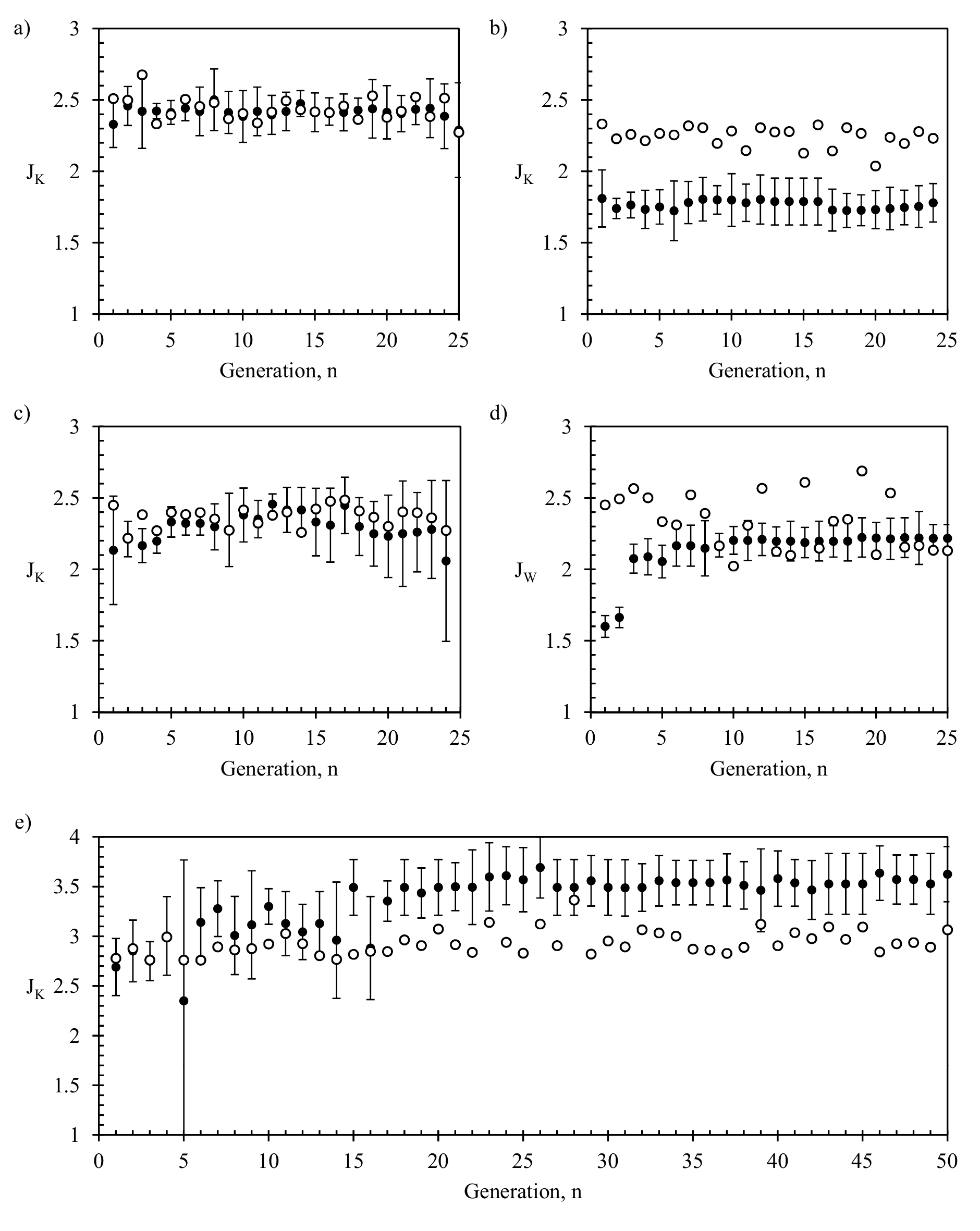}}% Images in 100% size
  \caption{Convergence of MLC process using various GP and experimental parameters: a) MaxK experiment with standard GP settings (see table~\ref{tab:standard_params}), b) MaxK experiment using a small population size ($N^{(n)}=100$, for  $n=1,\cdots,25$), c) MaxK experiment using a medium population size ($N^{(n)}=500$, for $n=1,\cdots,25$), d) MaxW experiment using a very small population size and large mutation rate ($N^{(n)}=50$, for $n=1,\cdots,25$; $P_r=0.0$, $P_x=0.35$, $P_m=0.65$), and e) MaxK experiment using the high-gain actuation amplitude controller ($N^{(n)}=500$, for $n=1,\cdots,50$).}
\label{fig:gp_param}
\end{figure}

The various effects of different GP parameters are sampled in figure~\ref{fig:gp_param}. Each of the diagrams in this figure shows values of cost function $J_K$ or $J_W$ for open-loop reference (white circles) and best individual (black circles) for each generation. Best MLC individual data points are averaged values of the objective function for each instance this particular individual has been tested. This means that if an individual appeared in several generations (regardless of whether it was the best in every generation) all of its evaluations contribute to the value, similar to what is used to calculate the cost function in MLC as presented in section~\ref{sec:exp_mlc}. The associated error bars denote the standard dispersion (assuming a Gaussian distribution) equal to three times the root-mean-square ({\it rms}) value of the best individual. Open-loop data points have no error bars since they are only evaluated once per generation. The dispersion of this data can be directly inferred from the diagrams. 

An MLC experiment using the standard parameter set (see table~\ref{tab:standard_params}) with a large initial and smaller subsequent population sizes is depicted in figure~\ref{fig:gp_param}(a). It is apparent that the initial generation yields already the best individual, after which the GP process is only confirming the solution. An alternative experiment is shown in figure~\ref{fig:gp_param}(b) where a sub-optimal solution is found in the first generation and it never evolves to a better one. This experiment used a small population size of 100 individuals in each generation, including the first. These results suggest that a Monte-Carlo random process \citep{metropolis1949monte}, in forming the first generation, can be used to obtain the best solution if the population is large enough. An experiment involving 500 individuals per generation is shown in figure~\ref{fig:gp_param}(c). Here, the best solution is also found in the first generation. This indicates that a population size of $N^{(1)}=500$ individuals is enough to yield a solution by random probing of the search space, in this experiment.

Figure~\ref{fig:gp_param}(d) shows the effect of a radical change in parameter settings. In this case, only 50 individuals were used in each generation, while genetic operations have been set to $0\,\%$ for replication, $65\,\%$ for mutation and $35\,\%$ for cross-over. Elitism is set to $N_e=5$ best individuals. These settings assured that the initial generation had distinctly sub-optimal individuals, but gave ample room for evolution by introducing much more mutation, \textit{i.e.} the ability to expand the search space beyond the possibilities of the first generation. A significant qualitative jump in cost function value of the best individual is apparent as early as the 3rd generation. This kind of behavior is a signature of a successful GP process. Testing time for this experiment until a stable best solution (which appears in generation $n=6$) is significantly shorter than any previously shown experiment.   

Another clearly successful GP process is shown in figure~\ref{fig:gp_param}(e), this time for the case of the high-gain amplitude controller. In this case, as explained in section~\ref{sec:alt_pi}, the complexity of the controlled plant was increased by the response of the actuation amplitude control system. In terms of GP parameters, this experimental run was not different from the run shown in figure~\ref{fig:gp_param}(c), \textit{i.e.} it used 500 individuals in each generation and a standard set of other genetic operation parameters. Likewise, we can see that the best solution in the first generation was approximately equal to best open-loop control, as expected when a large initial population size is used. However, the MLC process evolved a better solution (appearing in generation $n=15$), which no single frequency periodic actuation could have provided.

\subsection{Statistical analysis of MLC performance}\label{sec:stats}

When the data from the GP process presented in the previous section is analyzed, it can be seen that both open- and closed-loop actuation has a level of standard dispersion from which it is sometimes difficult to reach a clear conclusion of which one is better. This deviation in performance results from the natural perturbations in the wind tunnel and in the actuator system.

A more complete way of making a comparison between open- and closed-loop control in this experiment is to average out all the results across the experiment. Resulting mean and {\it rms} cost function values ($\overline J_K$, $\overline J_W$ and $J_{K_{rms}}$, $J_{W_{rms}}$) are shown in table~\ref{tab:stats}, for all experiments discussed in section~\ref{sec:results}. The cost function values for the open-loop reference are averaged across their single evaluations in every generation. The best MLC individual is chosen from the last generation, but the averaging is performed using each evaluation of this individual, from every generation in which it appeared previously. This procedure should be sufficient to ensure comparable statistical sample sizes.

\begin{table}
  \begin{center}
\def~{\hphantom{000}}
  \begin{tabular}{lcccccc}
		\multicolumn{1}{c}{} &
		\multicolumn{1}{c}{\phantom{00}} &
        \multicolumn{2}{c}{$x=200\,[\si{\milli\meter}]$} &
        \multicolumn{1}{c}{\phantom{00}} &	
        \multicolumn{2}{c}{$x=500\,[\si{\milli\meter}]$} \\
					\hline
%					&					&					&					&				&					&		\\	
	MaxW (LS)		&	\phantom{00}	& $\overline{J}_W$	&	$J_{W_{rms}}$	& \phantom{00} & $\overline{J}_W$	& $J_{W_{rms}}$ \\[3pt]
       Open-loop    &	\phantom{00}	& 2.32				&	0.20      		& \phantom{00} & 1.62				& 0.09 			\\
       MLC          &	\phantom{00}	& 2.20				&	0.08      		& \phantom{00} & 1.55				& 0.18 			\\
       				\hline	
%					& 					&					&					&				&					&				 \\
	MaxK (LS)		& 					& $\overline{J}_K$	& $J_{K_{rms}}$		& 				&$\overline{J}_K$	& $J_{K_{rms}}$  \\[3pt]
       Open-loop    & 					& 2.44 				& 0.08 				& 				& 1.85 				& 0.19			 \\
       MLC  		& 					& 2.52 				& 0.25 				& 				& 1.73 				& 0.13			\\
					\hline	
%					& 					&					&					&				&					&				 \\
	MinK (LS)		& 					& $\overline{J}_K$ & $J_{K_{rms}}$ 		& 				&$\overline{J}_K$ 	& $J_{K_{rms}}$ \\[3pt]
       Open-loop    & 					&	0.70 			& 0.04 				& 				& 0.81 				& 0.05 				\\
       MLC  		& 					&	0.75 			& 0.04 				& 				& 1.04 				& 0.12 				\\
              		\hline	
%					& 					&					&					&				&					&				 \\
	MaxW (HS)		& 					& $\overline{J}_W$ 	& $J_{W_{rms}}$ 	& 				&	$\overline{J}_W$ & $J_{W_{rms}}$ \\[3pt]
       Open-loop    & 					& 1.86 				& 0.05 				& 				& - 				& - \\
       MLC			& 					&	1.75 			& 0.08 				& 				& - 				& - \\
					\hline
%					& 					&					&					&				&					&				 \\
	MaxK (high-gain PI)&				& $\overline{J}_K$ 	& $J_{K_{rms}}$ 	& 				&	$\overline{J}_K$ & $J_{K_{rms}}$ \\[3pt]
       Open-loop    & 					&	2.93 			& 0.12 				& 				& - 				& - \\
       MLC  		& 					&	3.48 			& 0.16 				& 				& - 				& -\\       
  \end{tabular}
  \caption{Mean ($\overline J$) and standard deviation ($J_{rms}$) cost function values of the best open-loop control and MLC individuals, for all the experiments presented in section \ref{sec:results}.}
  \label{tab:stats}
  \end{center}
\end{table}

The values in table~\ref{tab:stats} indicate the open-loop actuation as having around 5\% of an average performance advantage, in most cases. Analysis of the {\it rms} values shows they are of the order of  5--10\% of the mean value. Depending on the experiment, such a high standard deviation can appear in either open- or closed-loop control case. Most obvious examples being the open-loop control in the MaxW (LS) experiment, and MLC best individual in the MaxK (LS) experiment, both for $x=\SI{200}{\milli\meter}$. Thus, a direct performance comparison between open- and closed-loop control is rendered inconclusive, considering relatively large standard deviations of the cost function values. It is clear, however, that both control types have a significant impact on the mixing layer.

\subsection{Robustness of MLC performance}\label{sec:robust}

\begin{figure}
  \centerline{\includegraphics[width=1\textwidth]{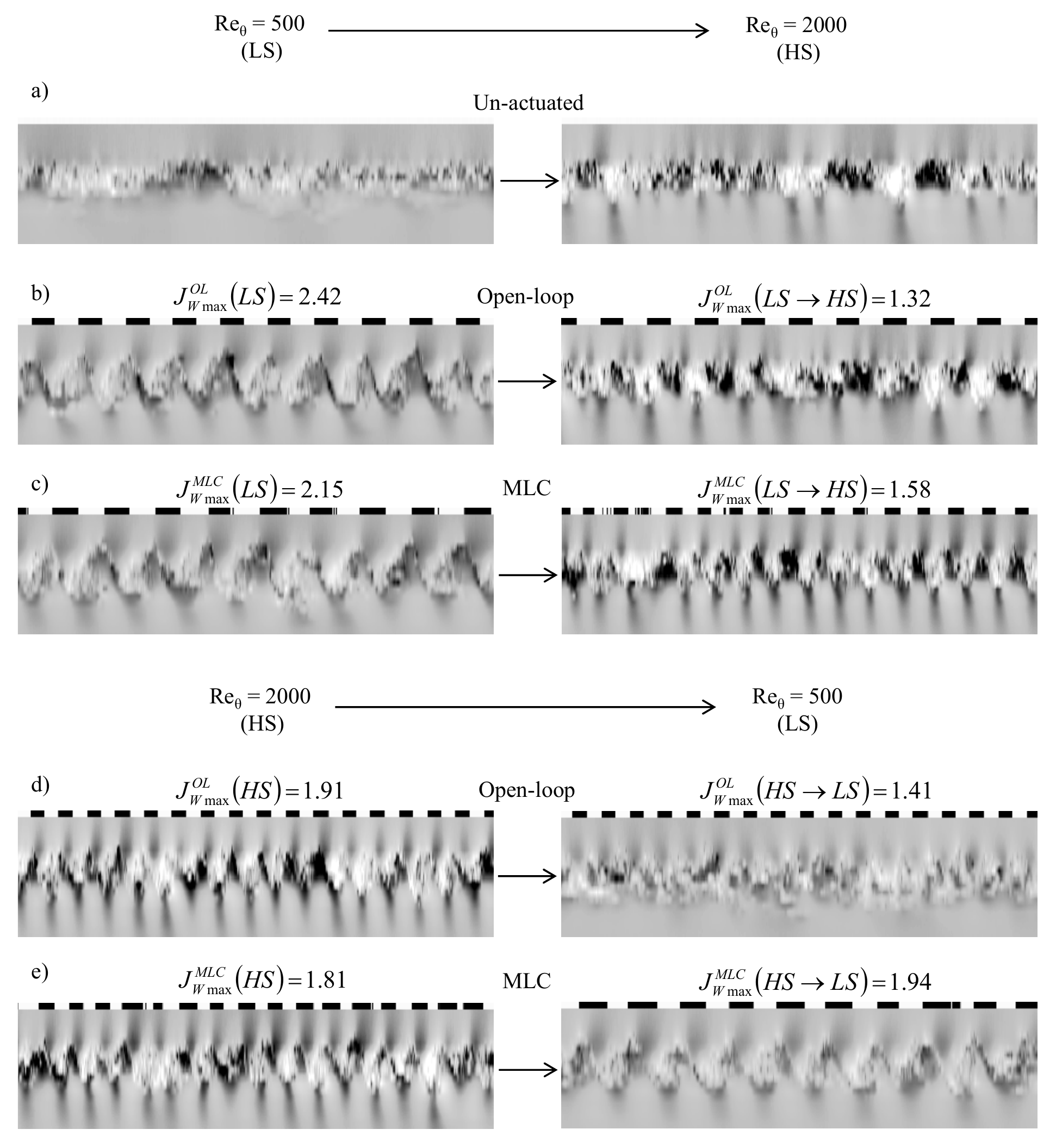}}% Images in 100% size
  \caption{Robustness tests of open- and closed-loop control, with respect to two different speed configurations of the mixing layer. The best open-loop and the best MLC individual are tested in the conditions for which they are optimal, and then re-evaluated in a different configuration. Test of increasing $\Rey_{\theta}$ corresponds to: a) un-actuated, b) optimal LS open-loop, and c) best MLC for LS. Second test, decreasing $\Rey_{\theta}$, is shown for: d) optimal open-loop for HS, and e) best MLC for HS. Pseudo-visualization is shown on a time scale of $t=\SI{0.5}{\second}$, but otherwise similarly as in figure~\ref{fig:200_wmax_res_500_wmax_res}.}
\label{fig:robust}
\end{figure}

A series of experiments was performed to explore the sensitivity of open- and closed-loop control to different flow conditions. Best scenarios of both types of control for maximization of objective function $W$ (see section~\ref{sec:max_w}) were tested in a simulated change of $\Rey_{\theta}$ of the mixing layer. The un-actuated flow is first evaluated in both LS ($\Rey_{\theta}=500$) and HS ($\Rey_{\theta}=2000$) conditions (shown in figure~\ref{fig:robust}a), for the purpose of estimating the results as a normalized cost function $J_W$.

In the first test of robustness (shown in figure~\ref{fig:robust}b and c), we consider LS as a starting condition. Optimal open-loop and MLC control, which were found for the LS configuration were tested, yielding $J_{\text{\tiny Wmax}}^{\text{\tiny OL}}(\text{LS})=2.42$ and $J_{\text{\tiny Wmax}}^{\text{\tiny MLC}}(\text{LS})=2.15$, respectively. Once these evaluations were complete, the wind-tunnel speed was changed to the HS settings and the controls were re-evaluated. The resulting cost function values are: $J_{\text{\tiny Wmax}}^{\text{\tiny OL}}(\text{LS\ding{213}HS})=1.32$ and $J_{\text{\tiny Wmax}}^{\text{\tiny MLC}}(\text{LS\ding{213}HS})=1.58$. 

The average total enhancement of the cost function, across the two flow conditions, can be computed as:

\begin{equation}
\langle J_{\text{\tiny Wmax}}\rangle_{\text{\tiny LS,HS}} = \frac{1}{2}[J_{\text{\tiny Wmax}}(\text{LS})+ J_{\text{\tiny Wmax}}(\text{LS\ding{213}HS})],
\label{eq:j_ls_hs}
\end{equation}

\noindent for both types of control. The average cost function yields $\langle J_{\text{\tiny Wmax}}^{\text{\tiny OL}}\rangle_{\text{\tiny LS,HS}}=1.87$ for open-loop and $\langle J_{\text{\tiny Wmax}}^{\text{\tiny MLC}}\rangle_{\text{\tiny LS,HS}}=1.865$ for MLC. We can conclude that when these two controls are engaged and a speed change occurs, both will yield a similar average enhancement of around 87\% of objective function $W$, compared to the un-actuated flow.

%The differences in the cost function values, between the two tests are: $\DeltaJ_W_{\text{\tiny OL}}^{\text{\tiny LS}}=1.1$ for open-loop, and $\DeltaJ_W_{\text{\tiny MLC}}^{\text{\tiny LS}}=0.57$ for MLC. 

%Their respective differences being: $\DeltaJ_W_{\text{\tiny OL}}^{\text{\tiny LS}}=0.5$, and $\DeltaJ_W_{\text{\tiny MLC}}^{\text{\tiny HS}}=0.13$. 

The second robustness test (shown in figure~\ref{fig:robust}d and e) is started with HS as the initial condition, followed by a test in the LS setting. Here we test open-loop and MLC controls which were found best in HS conditions. We note the cost function values: $J_{\text{\tiny Wmax}}^{\text{\tiny OL}}(\text{HS})=1.91$ and $J_{\text{\tiny Wmax}}^{\text{\tiny MLC}}(\text{HS})=1.81$ for the starting conditions, and $J_{\text{\tiny Wmax}}^{\text{\tiny OL}}(\text{HS\ding{213}LS})=1.41$ and $J_{\text{\tiny Wmax}}^{\text{\tiny MLC}}(\text{HS\ding{213}LS})=1.94$, for the changed conditions. The average cost function values in this case are: $\langle J_{\text{\tiny Wmax}}^{\text{\tiny OL}}\rangle_{\text{\tiny HS,LS}}=1.66$ for open-loop, and $\langle J_{\text{\tiny Wmax}}^{\text{\tiny MLC}}\rangle_{\text{\tiny HS,LS}}=1.875$ for MLC. In this reversed speed change test, MLC has kept its score of an average $W$ enhancement of 87\%, while open-loop yielded only 66\%. 

As seen in the previous section, open-loop performs slightly better than closed-loop when tested in the optimal conditions: $J_{\text{\tiny Wmax}}^{\text{\tiny OL}}(\text{LS})$ and $J_{\text{\tiny Wmax}}^{\text{\tiny OL}}(\text{HS})$ in figure~\ref{fig:robust}(b) and (d), respectively. When the conditions change, however, closed-loop control retains much of its effectiveness. The pseudo-visualization in figure~\ref{fig:robust}(c) shows that closed-loop $(\text{LS\ding{213}HS})$ adopts a global frequency of actuation very similar to the optimal open-loop in HS configuration, shown in figure~\ref{fig:robust}(d). A similar adaptation can be seen in figure~\ref{fig:robust}(e), where the best MLC individual in $\text{HS\ding{213}LS}$ configuration is very similar to the best open-loop actuation for LS conditions in figure~\ref{fig:robust}(b).

We can also compare $J_{\text{\tiny Wmax}}^{\text{\tiny OL}}(\text{LS\ding{213}HS})$ (figure~\ref{fig:robust}b) to $J_{\text{\tiny Wmax}}^{\text{\tiny OL}}(\text{HS})$ (figure~\ref{fig:robust}d). Non-optimal open-loop scores 60\% less $W$ enhancement compared to what an optimal actuation would achieve. In a reverse test, the comparison of $J_{\text{\tiny Wmax}}^{\text{\tiny OL}}(\text{HS\ding{213}LS})$ (figure~\ref{fig:robust}d) and $J_{\text{\tiny Wmax}}^{\text{\tiny OL}}(\text{LS})$ (figure~\ref{fig:robust}b) yields a 100\% reduction of effectiveness. MLC performs much better; in similar comparisons for both tests, only 21--23\% of reduction in effectiveness is recorded.

It can be concluded that open-loop can be a viable control method in varying flow conditions, only if it were to be applied using an adaptive control strategy (reference tracking or extremum seeking). Closed-loop control, on the other hand, shows an impressive ability to adapt and retain its effectiveness. Based on these results, a possible improvement of the MLC process could be to test multiple search spaces (based on flow conditions) in one experiment, in a search of an individual which scores the average best results across all conditions.

\subsection{Time cost of MLC experiments}\label{sec:time}

The study of the effects of parameter choice on the MLC convergence process reveals that very small population sizes are enough for the best solution to be obtained, and in just a few generations. These results have been repeated in both low- and high-speed mixing layer, and confirmed by several identical experimental runs. 

The experiment shown in figure~\ref{fig:gp_param}(d) indicates that the best solution can be seen as early as generation $n=6$. As mentioned before, this experiment was performed using population sizes of $N=50$ individuals in each generation. Consider that waiting for 2 more generations, until $n=8$, is necessary to confirm if the solution is converged; this means that the total experimental time for obtaining this solution, taking into account $\SI{20}{\second}$ for one evaluation, is $n \times N \times \SI{20}{\second} = \SI{2.2}{\hour}$. This is substantially faster than evaluating a large first generation of 500 individuals and waiting for two more generations in order to confirm the best solution, which would be the minimum required to obtain the best solution through a Monte-Carlo process ($3 \times 500 \times \SI{20}{\second} = \SI{8.3}{\hour}$). 

\begin{figure}
  \centerline{\includegraphics[width=1\textwidth]{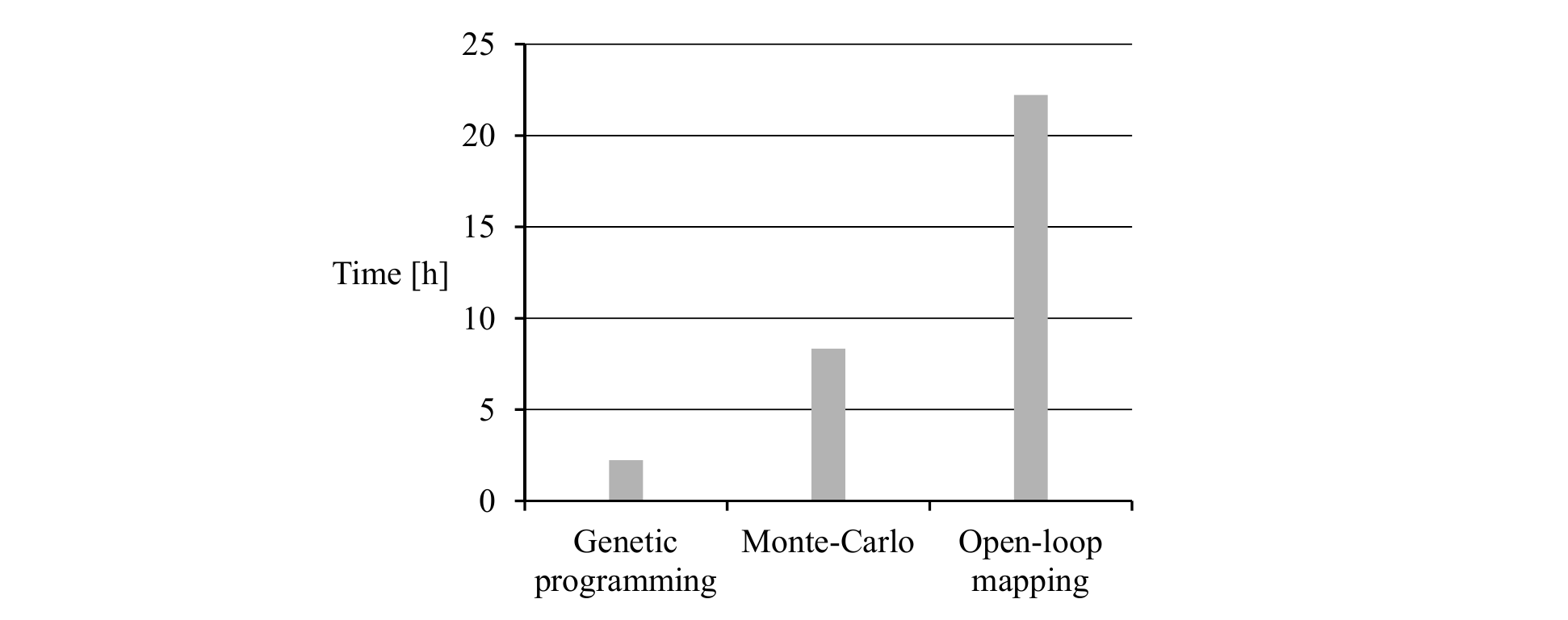}}% Images in 100% size
  \caption{Estimation of experimental time needed for obtaining a best actuation solution using genetic programming, Monte-Carlo random process and open-loop mapping.}
\label{fig:times}
\end{figure}

Both of these experiments are, in turn, much shorter than a full mapping of the frequency/duty cycle space of periodic forcing available to the actuator system. Mapping would imply testing of frequencies up to \SI{400}{\hertz} in steps of \SI{1}{\hertz} and repeating the process using ten different duty cycles ($400 \times 10 \times \SI{20}{\second} = \SI{22}{\hour}$). These projected time costs are comparatively shown in figure~\ref{fig:times}. If one adds testing in different flow configurations, the time costs would increase exponentially, putting the open-loop mapping at an even greater disadvantage. In this respect as well, MLC shows significant promise as a tool for exploring control of flow configurations whose dynamics are not well known in advance.

% =============== CONCLUSION =========================================%
\section{Conclusions and future directions}\label{sec:conclusion}

A model-free method for obtaining an optimal sensor-based control law for general multiple-input multiple-output (MIMO) plants, the \enquote{machine learning control} (MLC), is tested in an experimental mixing layer flow. MLC is based on genetic programming (GP), a function optimization method of machine learning. Randomly created linear or non-linear control laws are graded on-line with respect to a predefined cost function. Based on their success in minimizing or maximizing the cost function, the best are selected and evolved using genetic operations, to create the next generation of control law candidates. The process can be stopped when a satisfactory solution is found, or no further improvements are evident.

MLC has been successfully applied \citep{duriez_jfm_2014} to a closed-loop stabilization of a generalized two-frequency mean-field model, and closed-loop control for the maximization of the Lyapunov exponent (stretching) of the forced Lorenz equations. In the former case, MLC allows detection and exploitation of frequency cross-talk, in an unsupervised manner. By definition, frequency cross-talk is ignored in any linearized system. In the case of the forced Lorenz equations, MLC provides an increase of unpredictability which is a highly nonlinear phenomenon. MLC is also applied in an experiment featuring the backward facing step flow~\citep{gautier2014jfm}. Here, MLC yields a robust control law which causes a 50\% reduction of the recirculation bubble, using a flapping mode manipulation mechanism.

In the mixing layer experiment, we first present the results of mapping of open-loop forcing as a reference for evaluation of the closed-loop control. The best MLC control laws behave similar to the best open-loop forcing, with respect to frequency and duty cycle properties of the actuation signal. Since these control laws are sensor-based, they cannot be as perfectly regular as periodic forcing. Hence, an open-loop actuation, optimally selected for a given sensor location and an objective function, has a better average performance. The periodic forcing of a mixing layer has been thoroughly studied in experiments~\citep{Oster1982,Fiedler_1998_Cargese,Wiltse_GlezerJFM_1993}. These results all show that in order to change the local properties of a mixing layer, a clock-work type of actuation is sufficient. However, this clock-work needs to be precisely optimized for it to be effective at a given location in the convective flow of a mixing layer. This precise optimization is the weak point of open-loop actuation. By changing the initial Reynolds number of the mixing layer flow, the actuation parameters cease to be optimal and the impact of control is reduced. 

Closed-loop control, on the other hand, can be expected to be robust to such changes. The best MLC control laws succeed not only in distilling the optimal clock-work actuation from the sensors in the mixing layer flow, but also in showing a remarkable ability to adapt to the changes of the initial conditions. The best control laws, found for one set of initial conditions, adopted the optimal actuation parameters when tested in very different conditions, for which they were not designed. MLC-based closed-loop control retains 80\% of its effect for two extremes of the operational range of this wind tunnel. Compared to this, open-loop forcing is highly sensitive to such changes: its effect is reduced between 60\% to 100\% when conditions change.   

MLC has a large chance to detect and exploit otherwise invisible local extrema. Such a case has been presented in section~\ref{sec:alt_pi}, where MLC successfully exploited the non-linear behavior of the actuation system to amplify the fluctuation levels in the mixing layer. This result is unobtainable by any periodic forcing, but it comes as a consequence of including (unwittingly) the amplitude controller as a variable of the controlled plant. In this case, MLC proves to be a good detector of the boundaries of the targeted plant.

The success of the MLC search/optimization process depends on the choice of parameters. Results show that a good direction to adopt is to start with a small population size and use aggressive settings for mutation. This corresponds to a fast, low-resolution sweep of a large search space, while still retaining the ability to optimize a good solution through cross-overs. This process is also proven to be far superior in experimental time needed, compared to a simple open-loop parametric study. Such an experiment is not very time-consuming and, therefore, several runs could be easily performed to confirm whether a solution is a stable one. A non-converging solution of 4-5 repetitions of this experiment, would indicate that larger population sizes should be used and different genetic programming settings. However, starting with larger population sizes might prove to be unnecessary, if the problem turns out to be simple enough. A random creation of the first generation individuals may already find the best solution, through a Monte-Carlo random process. Nevertheless, while a single generation can yield a best solution in such a way, one cannot know that this is the best solution. Results from several more generations would be needed, so that a convergence of the solution becomes obvious.

\bigskip
All the results regarding the MLC convergence process and the closed-loop control performance may be summed up as follows:

\begin{itemize}
  \item Initial generation of random individuals can provide the best solution through a Monte-Carlo process, provided the problem at hand is simple enough and initial population size is large enough. In this case Genetic Programming can serve only to confirm the result or possibly optimize it.
  \item If the initial population does not yield the best solution, MLC, using a correct set of GP parameters, is able to achieve it in a very short time.
  \item For a given control problem, MLC will use any means provided to it to obtain a better solution. It is up to the user to provide a strong definition of the plant boundaries and minimize the possibility of external influences.
  \item Sensor-based control laws, using even the simplest forms of flow information from the mixing layer, can create an optimal actuation signal.
  \item Closed-loop control proves robust to changes in flow conditions. 
\end{itemize}

\bigskip

We foresee that MLC can be significantly improved
by adopting successful principles of control theory:
Firstly, the sensor signals may be filtered
so that they are less noise sensitive.
Secondly, the actuation command may be included as argument in the control law.
Thirdly, an ensemble of model-based control laws may constitute
the first generation to be improved by MLC.
Fourthly, the concept of reference tracking needs to be incorporated in future applications.
Fifthly, time can become are argument of the control law,
just like the sensor readings.
In this case, MLC would be able to find the best open-loop forcing
when closed-loop control is inevitably less effective.
In addition, spatial combinations of the sensors,
like in POD feedback control~\citep{Glauser2004,Parezanovic_FTC_2014} may also provide a means to mitigate both noise and sensitivity to sensor placement.

The model-free formulation makes MLC a very flexible method. It can be applied to any MIMO plant and use any cost function formulation. Theoretically, no {\it a priori} knowledge of the plant is needed, but an MLC experiment using "fast" settings can be very useful for obtaining quick additional information about the flow. This information can then be used to improve the initial guess with respect to the underlying physics. Not only could this lead to better optimization of the control laws, it could also potentially reduce the total time needed to understand the dynamics of the plant.   

% =============== ACKNOWLEDGEMENTS ===================================%
\begin{acknowledgments}
\section*{Acknowledgements}
The authors acknowledge the funding and excellent working conditions 
of the Senior Chair of Excellence
'Closed-loop control of turbulent shear flows 
using reduced-order models' (TUCOROM)
supported by the French Agence Nationale de la Recherche (ANR)
and hosted by Institute PPRIME. Additional funding by the ANR was provided via the SepaCoDe grant.
Marc Segond would like to acknowledge the support of the LINC
project (no. 289447) funded by EC's Marie-Curie ITN program
(FP7-PEOPLE-2011-ITN). 
We thank the Ambrosys Ltd. Society for Complex Systems Management, 
the Bernd Noack Cybernetics Foundation and Hermine Freienstein-Witt 
for additional support.

We appreciate valuable stimulating discussions
with the TUCOROM team 
(Jacques Bor\'ee, 
Eurika Kaiser, 
Nathan Kutz, 
Robert Niven, 
Michael Schlegel, 
Andreas Spohn 
and 
Gilles Tissot),
with the SepaCoDe team, in particular 
Azeddine Kourta and Michel Stanislas, 
our PMMH collaborators Jean-Luc Aider
and Nicolas Gautier, and Rudibert King, Christian Nayeri, Oliver Paschereit,
Rolf Radespiel, Peter Scholz and Richard Semaan.

%Jean-Paul Bonnet,
%Steven Brunton,
%Jo\"el Delville, 
%Thomas Duriez, 
%Nicolai Kamenzky, 
%Jacques Lewalle,
%Jean-Charles Laurentie,
%Marek Morzy\'nski,
%Michael Schlegel,
%Vladimir Parezanovic,
%and Gilles Tissot.
%\cek{Last but not least, we thank the referees
%for their thoughtful and helpful suggestions.}

Special thanks are due to Nadia Maamar 
for a wonderful job in hosting the TUCOROM visitors.
\end{acknowledgments}

% =============== BIBLIOGRAPHY =======================================%
\bibliographystyle{jfm}
\bibliography{biblio_jfm2014}

%%%%%%%%%%%%%%%%%%%%%%%%%%%%%%%%%%%%%%%%%%%%%%%%%%%%%%%%%%%%%%%%%%%%%%%
% APPENDIX
%%%%%%%%%%%%%%%%%%%%%%%%%%%%%%%%%%%%%%%%%%%%%%%%%%%%%%%%%%%%%%%%%%%%%%%%
\appendix   
\section{Linear System Identification and Models}\label{sec:era}

Here, we investigate linear models based on input--output data from the mixing layer experiment.  First, we introduce the particular methods for system identification employed for this data.  We then present results demonstrating the performance of linear models for predicting relevant quantities, such as a local turbulent energy content.  It is demonstrated that linear models are unable to capture significant flow features of the strongly nonlinear mixing layer.

\subsection{ERA/OKID approach}

To provide context for the unsupervised machine learning methods in this work, we first investigate the performance of linear models of the strongly nonlinear mixing layer.  In particular, we construct linear input--output models from impulse response data using the eigensystem realization algorithm (ERA) of \cite{ERA:1985}. ERA models have recently been shown to be equivalent to those obtained using balanced proper orthogonal decomposition (BPOD)~\citep{rowley:05pod,ERA:2009}, but without the need for adjoint simulations.  Therefore, ERA models of a given reduced order $r$  will capture the $r$ most observable and controllable fluid states in a balanced representation.  Balanced models are particularly useful for control since they faithfully capture the input--output relationship for a particular flow configuration and set of sensors and actuators~\citep{rowley:05pod,ilak:2008,bagheri:2009,illingworth:2010}.  Upper and lower error bounds on model performance are available in terms of the Hankel singular values when ERA/BPOD is used to reduce the model order of a linear system~\citep{rowley:05pod,dp:book}; this is due to a connection between ERA, BPOD, and balanced truncation.  Although we use ERA, there are many alternative methods to identify state-space models from input--output data~\citep{ljung:book}.

ERA is considered a {\em system identification} technique, since it is based on measurement data and does not rely on knowledge of an underlaying model.  Given output measurements for an impulse response in the actuation, ERA will produce a balanced state-space model of a desired model order $r$.  However, in many fluid experiments the impulse response data is noisy, and the underlying system may be strongly nonlinear.  In this case, we use frequency-rich input actuation sequences in conjunction with the observer/Kalman filter identification (OKID)~\citep{juang:1991}, which provides an cleaner estimate of the impulse response despite noisy measurements~\citep{Brunton:2012b,Brunton2014jfs}.  This filtered impulse response is then passed through ERA to obtain a reduced-order model.

\begin{figure}
  \centerline{\includegraphics[width=1\textwidth]{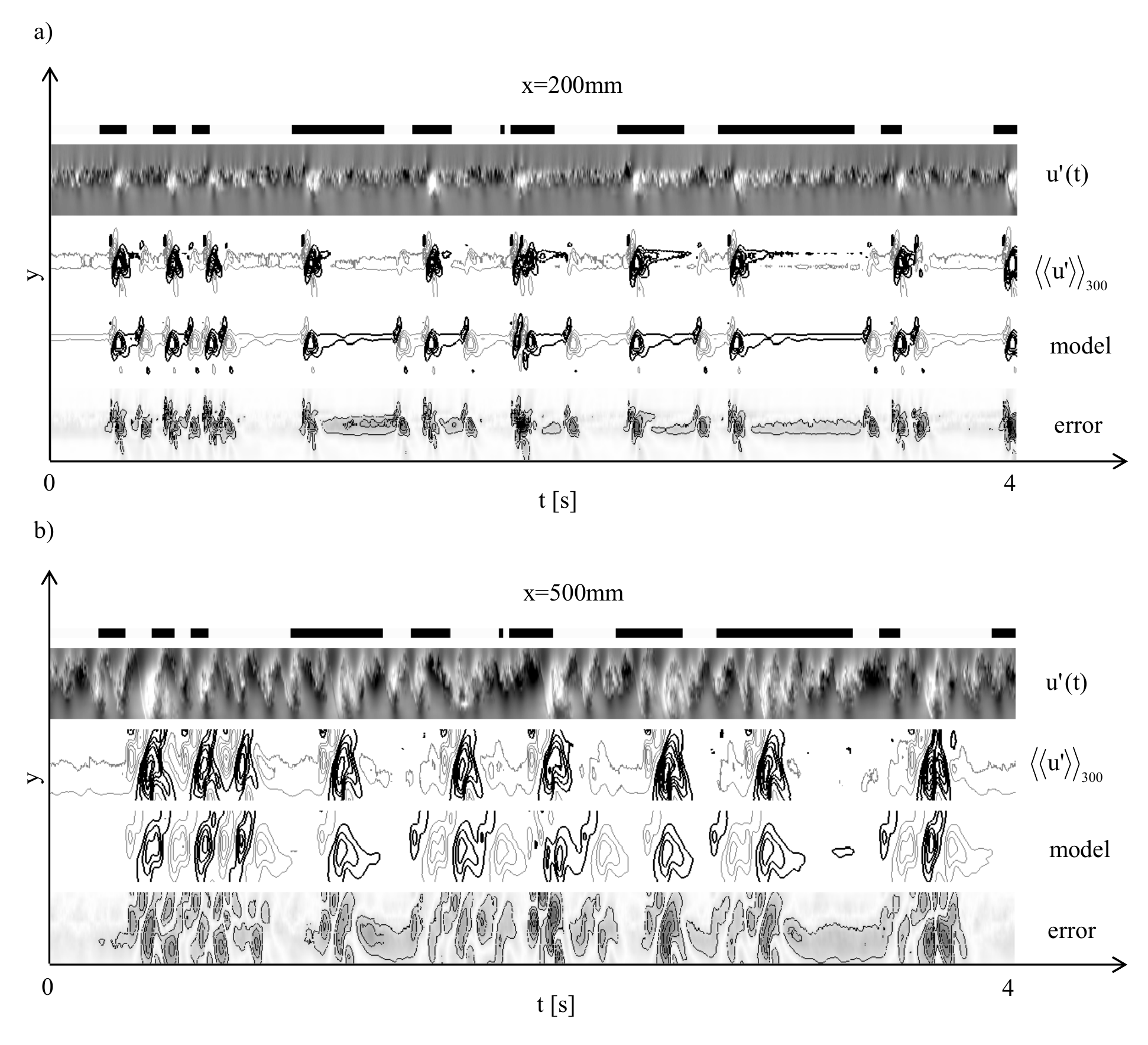}}% Images in 100% size
  \caption{Pseudo-visualization of the first four seconds of (top-down): the actuation signal, $u'(t)$, $\langle\langle u' \rangle\rangle_{300}$, model reconstruction and error, for a) $x=200\,mm$ and b) $x=500\,mm$.}
\label{fig:lin_model}
\end{figure}

\subsection{Experimental model ID}\label{model}

For the mixing layer experiment, we implement a pseudo-random actuation sequence where the jets are turned on and off for random intervals $\tau$ over a 30 second interval.  The variable $\tau$ is sampled from a Poisson distribution with $\lambda=4$ and a mean hold-time of $\tau=0.1$ seconds.  Hot-wire measurements (see section~\ref{sec:setup}) are collected for 300 repeated experimental runs and phase averaged, with respect to the actuation signal, to reduce irregular transient events. The idea of using a pseudo-random pulsed-blowing signal is inspired by~\cite{Kerstens:2011}.  

Before passing the phase-averaged data through OKID, it is first coarsened temporally to $\Delta t=0.004\,s$.  Next, since the downstream mixing layer is in a steady-state of saturated nonlinearity, corresponding to broad-band oscillations, the OKID filtered impulse response is multiplied by a hyperbolic tangent function that starts at $1$ and decays to $0$; this is critical to ensure a linearized impulse response that decays, resulting in stable models.  We then generate ERA models of order $r=15$ for each of the phase-averaged hot-wire signals; $r=15$ provides a balance between accuracy and model order, and results are qualitatively similar for $r=10$ and $r=20$.  

Figure~\ref{fig:lin_model} shows the actuation signal, an instance of the fluctuations $u'(t)$, the phase averaged signal $\langle\langle u' \rangle\rangle_{300}$, ERA model reconstruction and error for locations at $x=200\,mm$ and $x=500\,mm$.  From this figure, it is clear that the downstream velocity measurements are strongly correlated with the actuation signal, and that velocity fluctuation is enhanced.  Comparing the phase-averaged measurements $\langle\langle u' \rangle\rangle_{300}$ to a single instance $u'$, it is apparent that phase-averaging filters out many important transient flow nonlinearities, resulting in significantly decreased fluctuation levels.  Therefore, even a perfect model reconstruction of $\langle\langle u' \rangle\rangle_{300}$ would only capture a small portion of the true instantaneous fluctuation energy.  Moreover, the model reconstruction, although qualitatively similar to the phase-averaged measurements, has significant error on the same order of magnitude as the phase-averaged measurements.  

\begin{figure}
    \begin{center}
        \begin{tabular}{lccc}
			\rotatebox{90}{\phantom{000000000}$x=\SI{200}{\milli\meter}$} & 
            \begin{overpic}[width=.475\textwidth]{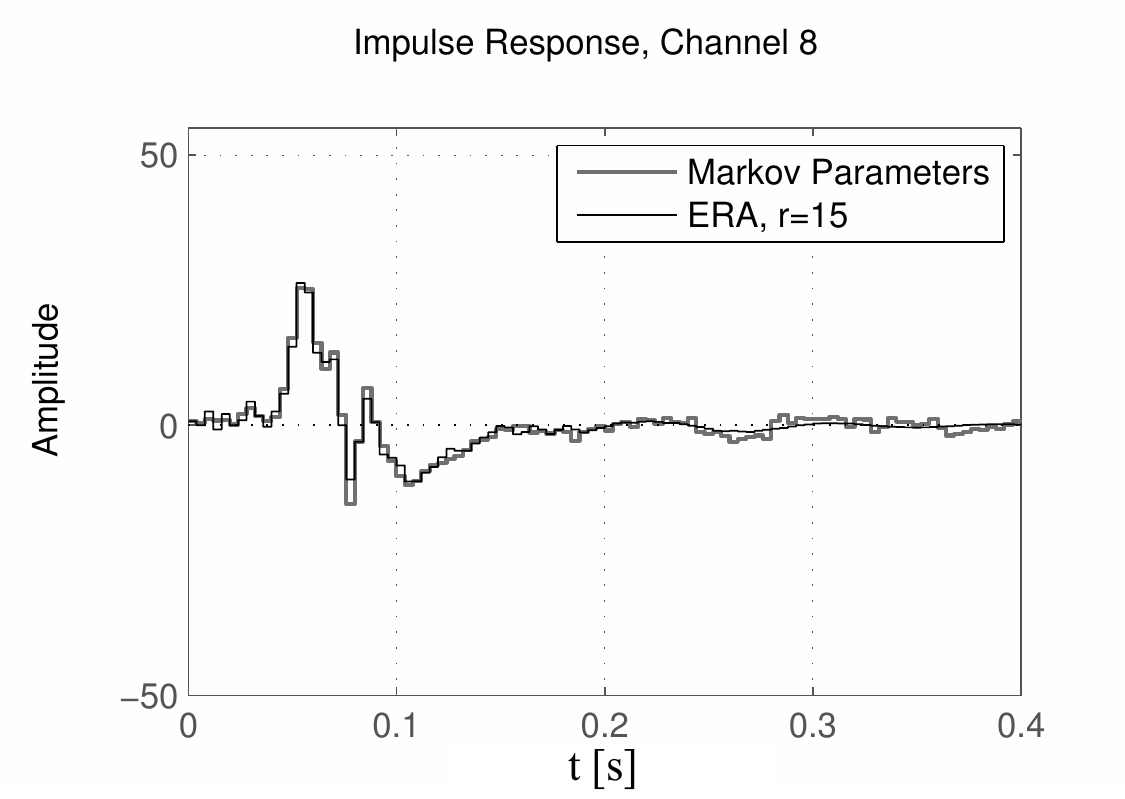}
            \end{overpic} &
            \begin{overpic}[width=.475\textwidth]{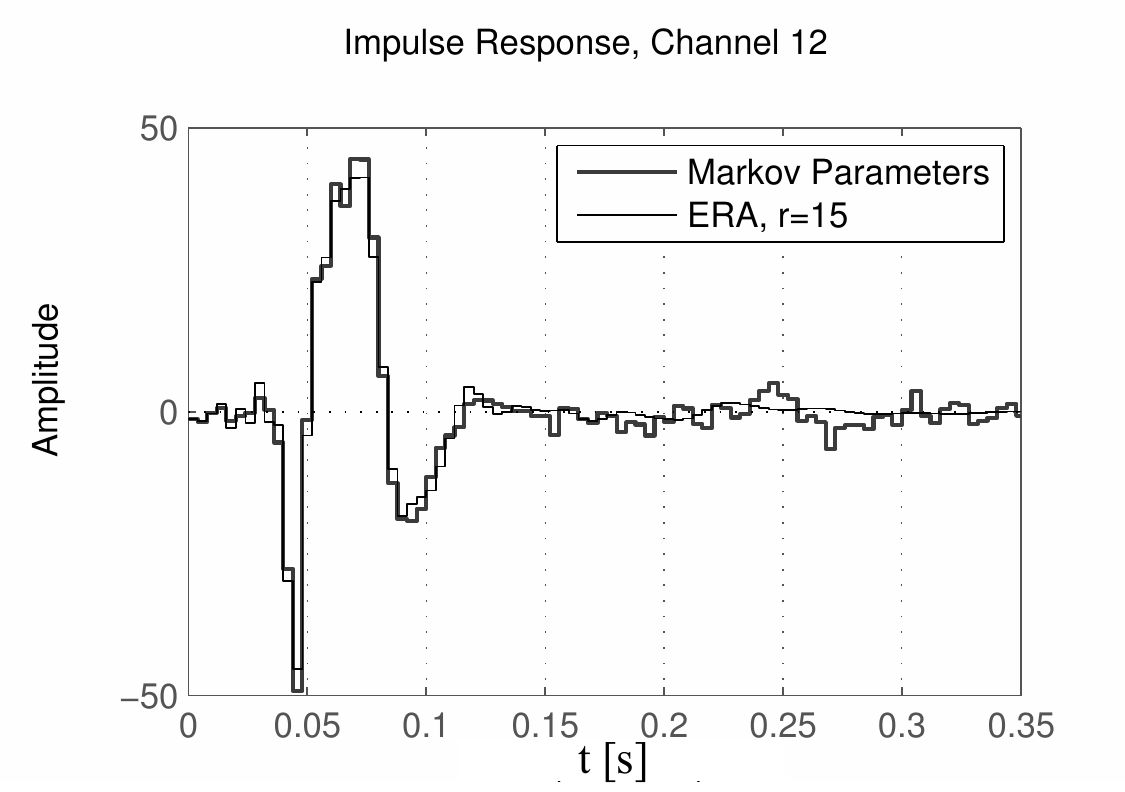}
            \end{overpic}\\
            & & \\
            \rotatebox{90}{\phantom{000000000}$x=\SI{500}{\milli\meter}$} &
			\begin{overpic}[width=.475\textwidth]{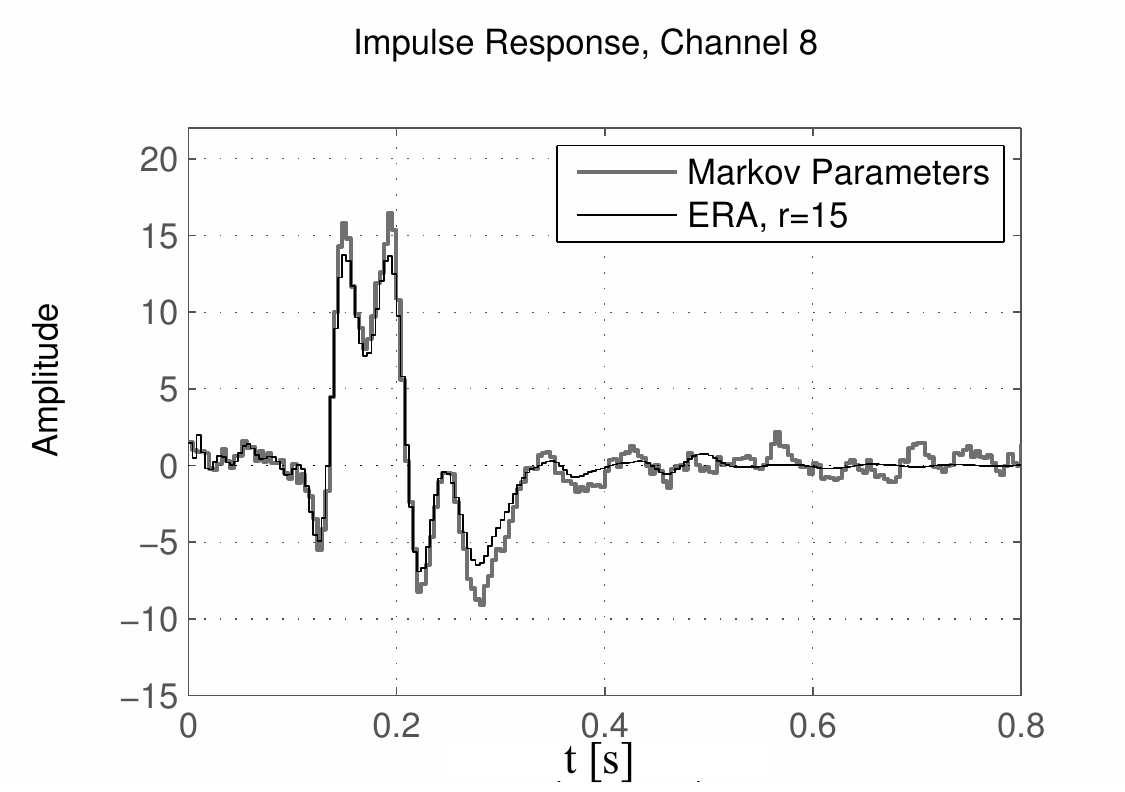}
            \end{overpic} &
            \begin{overpic}[width=.475\textwidth]{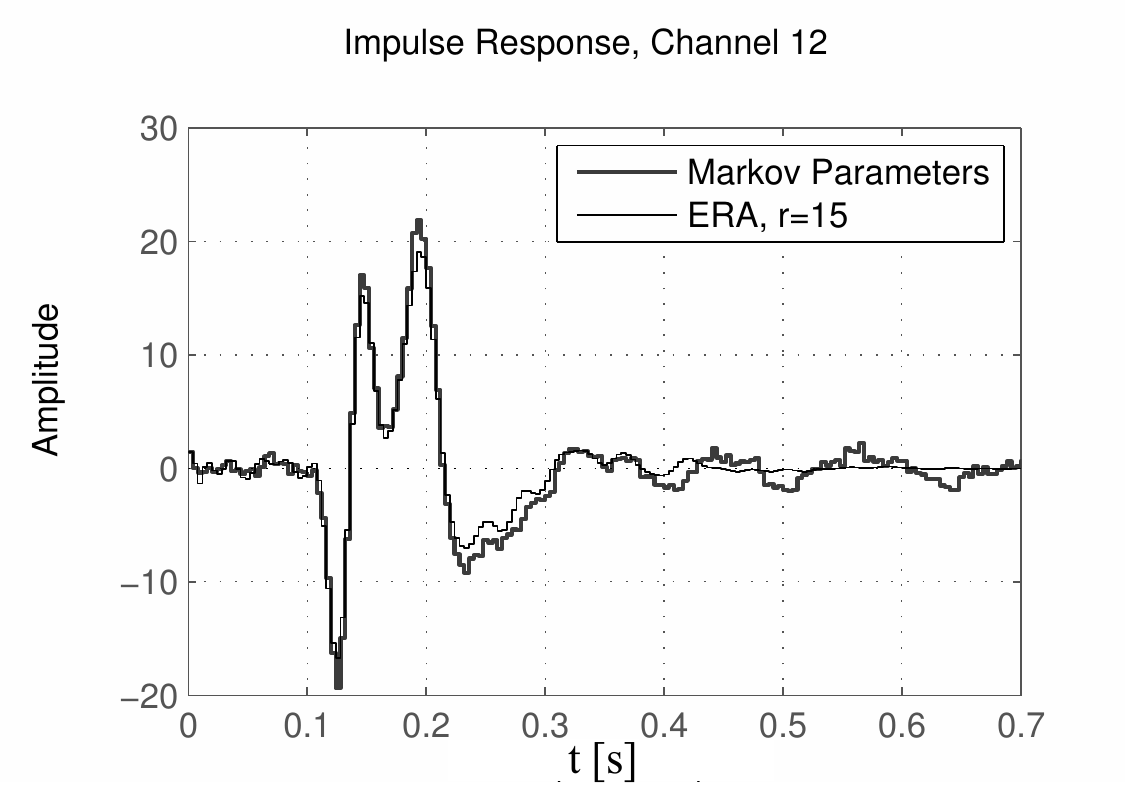}
            \end{overpic}\\ 
            
        \end{tabular}
        \vspace{-.05in}
        \caption{Markov parameters from OKID and impulse response of ERA model.}
        \label{fig:Markov}
    \end{center}
\end{figure}

Figure~\ref{fig:Markov}(top) shows the Markov parameters (output of OKID) and the ERA reconstruction for hot-wire channels 8 and 12 at $x=\SI{200}{\milli\meter}$.  Although the Markov parameters are well approximated by the ERA model, there is still significant reconstruction error when these models are used to approximate the original phase-averaged data, as shown in figure~\ref{fig:lin_model}.  This suggests that even the phase-averaged measurements exhibits strongly nonlinear responses to actuation.  Indeed, there are two important types of nonlinearity for this system.  The first type of transient, irregular flow events are removed by the phase-averaging procedure.  The second type of nonlinearity consists of regular, repeatable responses to actuation, which persist in the phase-averaged measurements.  The linear ERA models are incapable of capturing these nonlinearities, and therefore only capture a small portion of the phase-averaged response.  These two losses greatly diminish the effectiveness of linear models for predicting the nonlinear flow in response to actuation.  Similar plots of the Markov parameters, reconstruction, and error are shown for $x=500\,mm$ in figure~\ref{fig:Markov}(bottom) and figure~\ref{fig:lin_model}.

\begin{figure}
   \begin{center}
       \includegraphics[width=\textwidth]{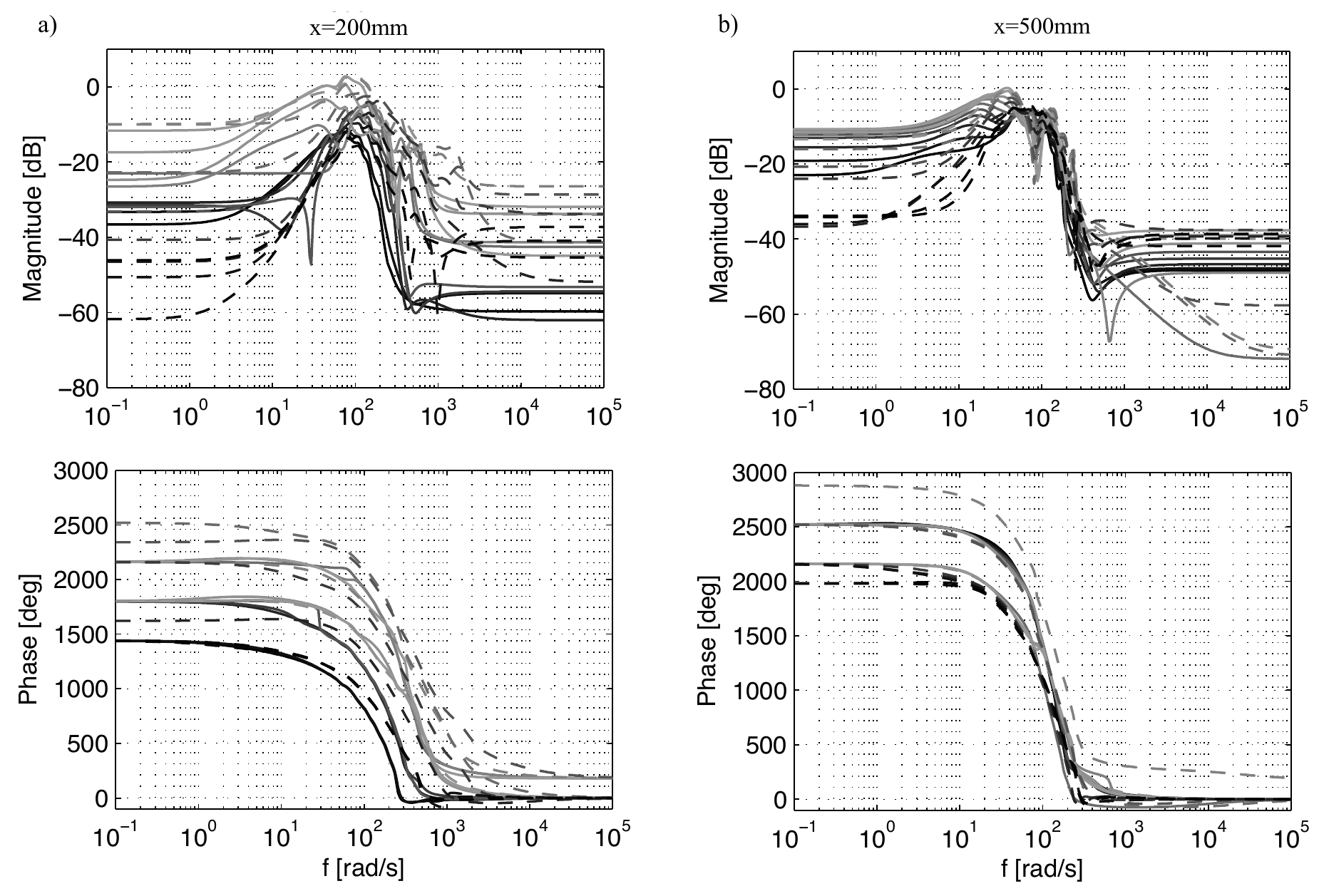}
        \caption{Bode plot for all hot-wire signals at a) $x=200\,mm$ and b) $x=500\,mm$.  Hot-wire channels $1$--$10$ are indicated by solid lines, and $11$--$19$ are dashed.  The lines are darker towards the middle of the rake and lighter toward the edges.}
        \label{fig.sysid.Bode}
    \end{center}
\end{figure}

% \begin{figure}
%     \begin{center}
%         \begin{overpic}[width=0.9\textwidth]{figures/long_200mm/f_fig05_bodeALLa_r15}
%         \end{overpic}\\
%         \begin{overpic}[width=0.9\textwidth]{figures/long_200mm/f_fig05_bodeALLb_r15}
%         \end{overpic}
%         \caption{Bode plot for all hot-wire signals at $x=200\,mm$.}
%         \label{fig.sysid.longBode200}
% \vspace{0.25in}
%    \end{center}
% \end{figure}

% \begin{figure}
%    \begin{center}
%         \begin{overpic}[width=0.9\textwidth]{figures/long_500mm/f_fig05_bodeALLa_r15}
%         \end{overpic}\\
%         \begin{overpic}[width=0.9\textwidth]{figures/long_500mm/f_fig05_bodeALLb_r15}
%         \end{overpic}        
%         \caption{Bode plot for all hot-wire signals at $x=500\,mm$.}
%         \label{fig.sysid.longBode500}
%     \end{center}
% \end{figure}

\begin{figure}
  \centerline{\includegraphics[width=1\textwidth]{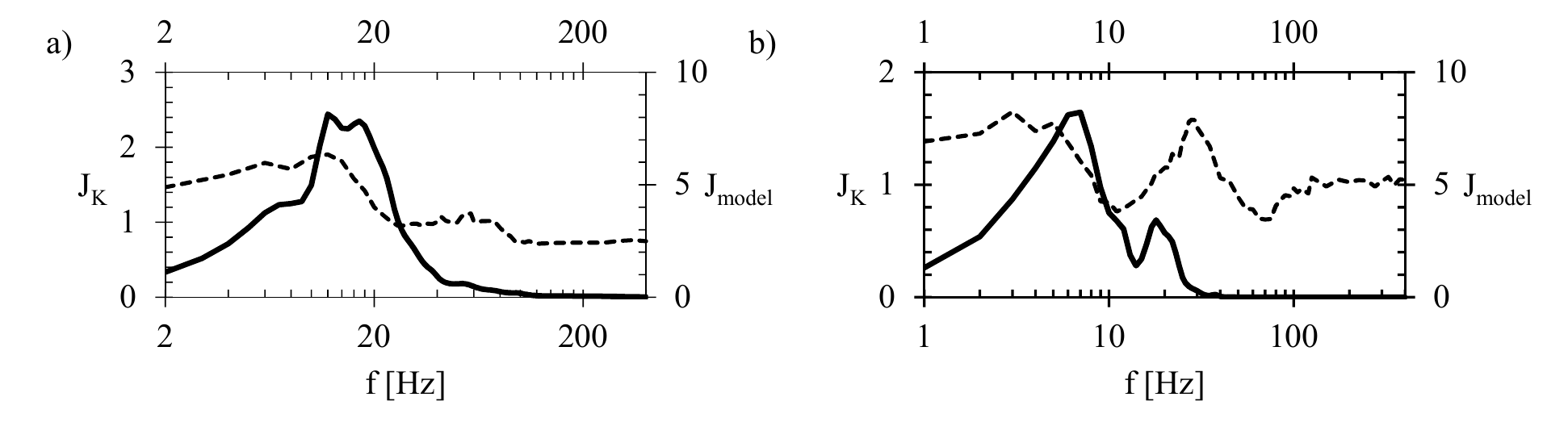}}% Images in 100% size
  \caption{Prediction of the mixing layer response to open-loop actuation according to the linear model (solid line), compared to the experiment (dashed line), for a) $x=200\,mm$ and b) $x=500\,mm$.}
\label{fig:model_j}
\end{figure}

The frequency responses (i.e., Bode plots) of the ERA models for all hot-wire channels are shown in figure~\ref{fig.sysid.Bode} for the $x=200\,mm$ and $x=500\,mm$ locations.  First, notice the presence of broadband resonant behavior in the middle-frequency range, as well as the presence of a small feed-through term, indicated by the high-frequency asymptote of the magnitude plot.  As the hot-wire rake is moved downstream, the feed-through term reduces significantly, as expected.  The resonant phenomenon also becomes more coordinated, and the phase plot indicates a larger time-delay.  From the magnitude plots, it is possible to estimate the energy-based cost function $J_K$ at various frequencies by summing the square of the magnitude across all hot-wires. The plots of this reconstruction of the cost function $J_\text{model}$ are shown in figure~\ref{fig:model_j} for the $x=200\,mm$ and $x=500\,mm$ locations.  These predictions correspond to the case of $50\%$ duty cycle actuation since both the ``on" and ``off" periods in the input actuation signal were sampled from the same Poisson distribution. Comparing with the experimental values from figure~\ref{fig:j_vs_f_200_500} at low frequencies, the linear model prediction of the cost function $J_\text{model}$ has similar features, but the amplitude of fluctuation energy is off by a factor of three. Moreover, the linear models incorrectly predict zero velocity fluctuation energy at high frequencies.

There are a number of important conclusions that may be drawn from this linear model identification.  First, it is clear that linear models obtained from the ERA/OKID procedure on phase-averaged hot-wire measurements have significant errors, on the same order as the averaged measurements.  The close agreement between Markov parameters and the ERA model indicates that this error is due to nonlinear flow responses that persist in the phase-averaged signal.  Moreover, the phase-averaged measurements fail to capture a significant portion of the turbulent fluctuations seen in a single instance of the response to actuation.  These compound errors highlight the severe limitations of linear systems to characterize the strongly nonlinear mixing layer.  However, the frequency response and predicted $J_\text{model}$ do capture some important qualitative features.

%%%%%%%%%%%%%%%%%%%%%%%%%%%%%%%%%%%%%%%%%%%%%%%%%%%%%%%%%%%%%%%%%%%%%%%
% FIN
%%%%%%%%%%%%%%%%%%%%%%%%%%%%%%%%%%%%%%%%%%%%%%%%%%%%%%%%%%%%%%%%%%%%%%%
\end{document}